\documentclass{jfm}

\usepackage{graphicx}
\usepackage{natbib}
\usepackage{amssymb}
\usepackage{amsmath}
\usepackage{dashrule}
\usepackage{color}
\usepackage{psfrag}
\usepackage{bm}
\usepackage{hyperref}
\hypersetup{
	colorlinks=true,
	linkcolor=black,
	citecolor=black
}
\usepackage{calrsfs}
\usepackage{tikz}
\usetikzlibrary{shapes,arrows}

\newcommand\Ray{\mbox{\textit{Ra}}}  
\newcommand\Nus{\mbox{\textit{Nu}}}  
\newcommand\Ric{\mbox{\textit{Ri}}}  
\newcommand{\R}{\mathcal{R}}

\newcommand\ie{i.e.\ }
\newcommand\cf{cf.\ }

\newcommand{\romd}{\mathrm{d}}

\shorttitle{Bulk scaling in VC}
\shortauthor{C. S. Ng, A. Ooi, D. Lohse and D. Chung}
\title{Bulk scaling in wall-bounded and homogeneous vertical natural convection}
\author
{Chong Shen Ng$^1$
	\thanks{Email address for correspondence: chongn@unimelb.edu.au},\ns
	Andrew Ooi$^1$, Detlef Lohse$^{2,3}$, and Daniel Chung$^1$}

\affiliation{$^1$Department of Mechanical Engineering,
	The University of Melbourne, Victoria 3010, Australia\\[\affilskip]
	$^2$Physics of Fluids Group, MESA+ Institute and J.\ M.\ Burgers Centre for 
	Fluid Dynamics and Max Planck Center Twente,\\University of Twente, P.O. Box 
	217, 7500AE Enschede, The Netherlands\\[\affilskip]$^3$Max Planck Institute for 
	Dynamics and Self-Organization, 37077 G\"{o}ttingen, Germany}

\date{?; revised ?; accepted ?. - To be entered by editorial office}

\begin{document}
\maketitle

\begin{abstract}

Previous numerical studies on homogeneous Rayleigh--B{\'e}nard convection, which is Rayleigh--B{\'e}nard convection (RBC) without walls, and therefore without boundary layers, have revealed a scaling regime that is consistent with theoretical predictions of bulk-dominated thermal convection. In this so-called asymptotic regime, previous studies have predicted that the Nusselt number ($\Nus$) and the Reynolds number ($\Rey$) vary with the Rayleigh number ($\Ray$) according to $\Nus\sim\Ray^{1/2}$ and $\Rey\sim\Ray^{1/2}$ at small Prandtl number ($\Pran$). In this study, we consider a flow that is similar to RBC but with the direction of temperature gradient perpendicular to gravity instead of parallel; we refer to this configuration as vertical natural convection (VC). Since the direction of the temperature gradient is different in VC, there is no exact relation for the average kinetic dissipation rate, which makes it necessary to explore alternative definitions for $\Nus$, $\Rey$ and $\Ray$ and to find physical arguments for closure, rather than making use of the exact relation between $\Nus$ and the dissipation rates as in RBC. Once we remove the walls from VC to obtain the homogeneous setup, we find that the aforementioned $1/2$-power-law scaling is present, similar to the case of homogeneous RBC. When focussing on the bulk, we find that the Nusselt and Reynolds numbers in the bulk of VC too exhibit the $1/2$-power-law scaling. These results suggest that the $1/2$-power-law scaling may even be found at lower Rayleigh numbers if the appropriate quantities in the turbulent bulk flow are employed for the definitions of $\Ray$, $\Rey$ and $\Nus$. From a stability perspective, at low- to moderate-$\Ray$, we find that the time-evolution of the Nusselt number for homogenous vertical natural convection is unsteady, which is consistent with the nature of the elevator modes reported in previous studies on homogeneous RBC.

\end{abstract}

\section{Introduction}

\begin{figure}
	\centering
	\centerline{\includegraphics{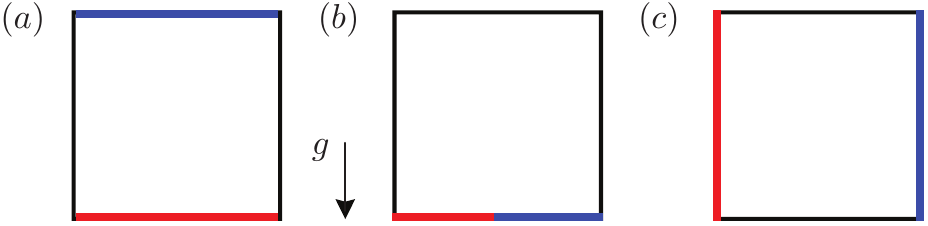}}
	\caption{\label{fig:3configs}Examples of wall-bounded configurations for ($a$) 
	Rayleigh--B{\'e}nard convection (RBC) ($b$) horizontal convection (HC) and ($c$) 
	vertical natural convection (VC). For illustration purposes, the configurations 
	are drawn in a cavity with aspect ratio equal to one. Heated walls are indicated 
	in red and cooled walls in blue. $g$ is the gravity vector.}
\end{figure}

Thermally driven flows play a crucial role in nature and are associated with many 
engineering flows. To study such flows, researchers typically consider idealised 
setups which include (figure \ref{fig:3configs}\textit{a}) the classical 
Rayleigh--B{\'e}nard convection, or RBC 
\citep{Ahlers+others.2009,Lohse+Xia.2010,Chilla+Schumacher.2012}, where fluid is 
confined between a heated bottom plate and a cooled upper plate, (figure 
\ref{fig:3configs}\textit{b}) horizontal convection, or HC
\citep{Hughes+Griffiths.2008,Shishkina+Grossmann+Lohse.2016}, where fluid is heated 
at one part of the bottom plate and cooled at some other part, and (figure 
\ref{fig:3configs}\textit{c}) vertical natural convection, or VC 
\citep{Ng+Ooi+Lohse+Chung.2014,Ng+Ooi+Lohse+Chung.2017}, where fluid is confined 
between two vertical walls, one heated and one cooled, \ie the flow is driven by a 
horizontal average heat flux. Alternative configurations of the VC flow, such as in 
a confined cavity \citep{Patterson+Armfield.1990,Yu+Li+Ecke.2007} and in a confined 
cylinder \citep{Shishkina+Horn.2016,Shishkina.2016.momentum} have also been 
investigated. In all these studies on thermal convection, there is a 
common interest to physically understand and predict how the temperature difference 
imposed on the flow (characterised by the Rayleigh number $\Ray$) influences the 
heat flux (characterised by the Nusselt number $\Nus$) and the degree of turbulence 
(characterised by the Reynolds number $\Rey$). With such relations, one is able to 
avoid relying on empirical relationships that are undetermined outside the range of 
calibration and which ignore the underlying physics.

At high $\Ray$, \cite{Kraichnan.1962} and 
\cite{Grossmann+Lohse.2000,Grossmann+Lohse.2001,Grossmann+Lohse.2002,Grossmann+Lohse.2004}
 -- hereinafter referred to as the GL theory -- predicted the so-called asymptotic ultimate-regime scaling where 
\begin{subeqnarray}\label{eqn:HalfPowerLaws}
	\gdef\thesubequation{\theequation \mbox{\textit{a},\textit{b}}}
	\Nus \sim \Ray^{1/2}, \quad	\Rey \sim \Ray^{1/2},
\end{subeqnarray}
\returnthesubequation
for low $\Pran$-values, for instance, when $\Pran \leqslant 1$.
(The $\Pran$-dependence of $\Nus\sim\Pran^{1/2}$ and 
$\Rey\sim\Pran^{-1/2}$ predicted by the GL theory for this asymptotic ultimate regime was confirmed in \cite{Calzavarini+others.2005} in the case of homogeneous RBC. For 
homogeneous VC, the $\Pran$-dependence is expected to be the same, but is beyond the 
scope of this paper.) These scaling relations contain logarithmic corrections when 
boundary layers or plumes are prominent \citep{Grossmann+Lohse.2011}. In numerical 
studies that seek to model only bulk thermal convection, \ie without boundary 
layers, the $1/2$-power-law scalings were indeed subsequently reported: 
\cite{Lohse+Toschi.2003} and \cite{Calzavarini+others.2005,Calzavarini+others.2006} 
discounted the influence of boundary layers by simulating a triply periodic 
configuration for RBC, termed homogeneous RBC. \cite{Schmidt+others.2012} 
numerically studied the same flow but with lateral (no-slip) confinement and also 
reported the $1/2$-power-law scaling
despite the presence of lateral boundary layers. In experiments, bulk convection is 
mimicked by measuring plate-free convection, \ie in a vertical channel connecting a 
hot chamber at the bottom and a cold chamber at the top 
\citep{Gibert+others.2006,Gibert+others.2009,Tisserand+Others.2010}, or by measuring 
fluid mixing in a long vertical pipe \citep{Cholemari+Arakeri.2009}. The use of 
alternative length scales to recover the $1/2$-power scaling in the bulk-dominated
flow has also been suggested \citep{Gibert+others.2006,Gibert+others.2009}. Later,
\cite{Riedinger+Tisserand+others.2013} conducted experiments by tilting the vertical
channel and in doing so introduced a gravitational component which is orthogonal
to the axis of the channel. The study reported a $1/2$-power slope at high $\Ray$ 
and low channel tilt angle (relative to the vertical). Recently, 
\cite{Frick+others.2015} investigated the effect of tilting in a sodium-filled
cylinder of aspect ratio equal to $5$ with a heated lower plate and a cooled upper 
plate. The study found that the heat transfer is more effective when the cylinder is 
tilted by $45^\circ$ compared to when the cylinder is tilted by $0^\circ$ and 
$90^\circ$. \cite{Shishkina+Horn.2016} obtained similar results in their numerical 
study where they compared RBC and VC by gradually tilting a fully enclosed 
cylindrical vessel.	Using the same cylindrical vessel, 
\cite{Shishkina.2016.momentum} then found that the effective power-law 
scaling in VC is smaller than $1/2$ because of geometrical confinement.

\begin{figure}
	\centering
	\centerline{\includegraphics{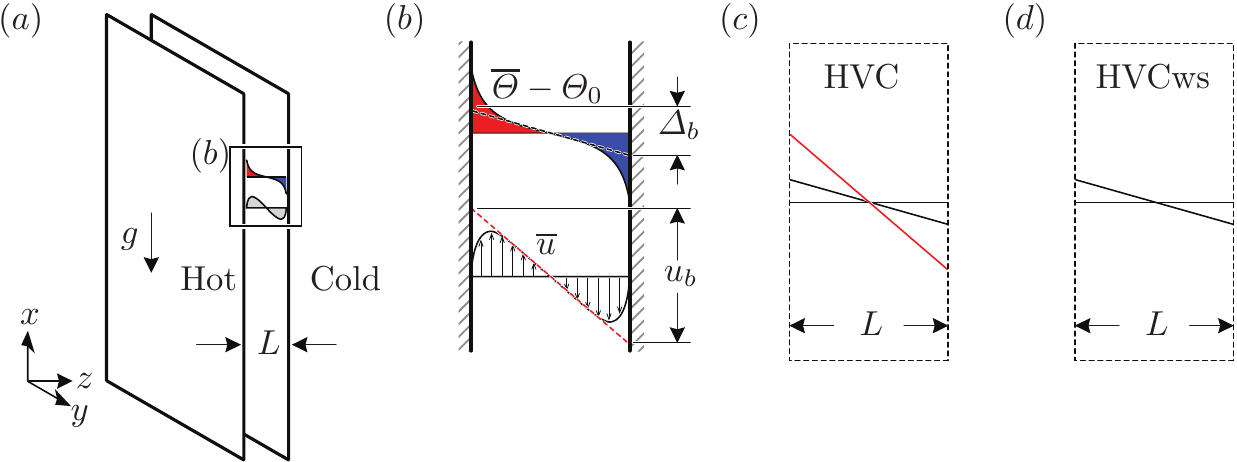}}
	\caption{\label{fig:VCSetup}($a$) Setup of VC. ($b$) Illustration of the mean temperature profile (top) and mean streamwise velocity profile (bottom). Both mean profiles are statistically antisymmetric about the channel centreline. 	($c$) Setup of HVC with a mean temperature and velocity gradient. ($d$) Setup of HVCws with only a mean temperature gradient. The black slopes in ($c$) and ($d$) represent the temperature gradient $\Delta_b/L$, whereas the red slope in 	($c$) represents the uniform mean shear $u_b/L$. Note that in ($a$), 	periodicity is applied in $x$- and $y$-directions only, whereas in ($c$) and 	($d$), periodicity is applied in all three directions, illustrated by the dashed boundaries.}
\end{figure}

In the present study, we investigate the scaling relations of VC in a triply periodic domain (figure \ref{fig:VCSetup}$c$ and $d$) using an approach that is similar to previous studies on bulk scaling for RBC \citep[\eg][]{Lohse+Toschi.2003,Calzavarini+others.2005}. The numerical setup of this homogeneous flow is described in \S\,\ref{sec:ComputParam}. Our objective is to determine whether the asymptotic ultimate $1/2$-power-law scaling can similarly be found in an idealised setup of VC which is free of influences from the boundary layers. To achieve this, we adopted two approaches: (\textit{i}) homogeneous VC with a constant mean temperature and velocity gradient in the horizontal direction, which we denote as HVC (figure \ref{fig:VCSetup}$c$), and (\textit{ii}) homogeneous VC with only a constant horizontal mean temperature gradient and no velocity gradient (or shear), which we denote as HVCws (figure \ref{fig:VCSetup}$d$). Although the flow in (\textit{i}) closely resembles the characteristics of the bulk flow in VC, we emphasize that the homogeneous and wall-bounded setups are different because for the homogeneous case, the energy transport to the walls is absent. In addition, the flow in ($ii$) without shear is evidently fictitious since a velocity gradient is present in the bulk of VC. In other words, both homogeneous cases (\textit{i}) and (\textit{ii}) are simplifications of VC, simplifications that are amenable to the $1/2$-power-law scaling arguments as has been shown previously for homogeneous RBC \citep[\eg][]{Lohse+Toschi.2003}.

The paper is structured as follows: We begin by describing the respective dynamical equations for HVC and HVCws in \S\,\ref{sec:DynamicalEquations}, which are numerically solved. Using scaling arguments, we relate both HVC and HVCws with the asymptotic ultimate $1/2$-power-law scaling in \S\,\ref{sec:UltimateScalingInHVC}. In \S\,\ref{sec:ComputParam}, we outline the numerical setups and direct numerical simulation (DNS) datasets, the latter of which is used to test the assumptions employed in our scaling arguments. Then, to assist comparisons with VC, we compare the dynamical lengthscales of HVC and HVCws with VC in \S\,\ref{sec:CompareStatistics}. In \S\,\ref{sec:ScalingRelations}, we compute the scaling of turbulent quantities of the homogeneous setups and show that $\Nus$ and $\Rey$ appear to follow the $1/2$-power-law scaling, just as in homogeneous RBC and consistent with the theoretical predictions by \cite{Kraichnan.1962}, the GL-theory and the scaling arguments in \S\,\ref{sec:UltimateScalingInHVC}. Inspired by the scaling of the turbulent quantities, we apply the insight gained to VC and find that the scaling of the turbulent bulk quantities in VC also exhibit the $1/2$-power-law scaling. Finally, in \S\,\ref{sec:ExponentialGrowth}, we compare the stability of the solutions for HVCws and homogeneous RBC, the latter of which is known to exhibit unstable, so-called `elevator modes' at low Rayleigh numbers. Such modes are associated with exponentially growing values of $\Nus$ followed by sudden break-downs \citep{Calzavarini+others.2005,Calzavarini+others.2006} and have been reported in similar studies, such as in laterally confined and axially homogeneous RBC \citep{Schmidt+others.2012}. In \S\,\ref{sec:CompareToDNS}, when we compared the stability analyses to data from our DNS of HVCws, we find that the unsteady solutions are also present in HVCws at low $\Ray$, but nonetheless both $\Nus$ and $\Rey$ appear to follow the $1/2$-power-law scaling.

\section{Dynamical equations} \label{sec:DynamicalEquations}
We begin with the general form of the governing equations for VC, where we 
invoke the Boussinesq approximation so that the density 
fluctuations are considered small relative to the mean. The governing continuity, 
momentum and energy equations for the velocity field 
$u_i(x_i,t)$ and the temperature field $\varTheta(x_i,t)$ are respectively given by,
\begin{subeqnarray} \label{eqn:GoverningEquations}
	\dfrac{\partial u_j}{\partial x_j} &=& 0, \\
	\dfrac{\partial u_i}{\partial t} + \dfrac{\partial u_j u_i}{\partial x_j} &=&
	-\dfrac{1}{\rho_0}\dfrac{\partial p}{\partial x_i} 
	+ \nu\dfrac{\partial^2 u_i}{\partial x_j^2} 
	+ g\beta (\varTheta-\varTheta_0)\delta_{i1}, \\
	\dfrac{\partial \varTheta}{\partial t} 
	+ \dfrac{\partial u_i\varTheta}{\partial x_i} &=&
	\kappa\dfrac{\partial^2\varTheta}{\partial x_i^2}.
\end{subeqnarray}
\returnthesubequation
The coordinate system $x$, $y$ and $z$ (or $x_1$, $x_2$ and $x_3$) refers to the vertical streamwise direction that is opposite to gravity, spanwise and wall-normal directions, respectively. The pressure field is denoted by $p(x_i,t)$. For VC (see figure \ref{fig:VCSetup}$a$), we define $\varTheta_0 = (T_h + T_c)/2$ as the reference temperature, $\Delta \equiv T_h - T_c$ the temperature difference between the two walls, which are separated by a distance $L$, and $g$ as the gravitational acceleration. For the fluid, we specify $\beta$ as the coefficient of thermal expansion, $\nu$ as the kinematic viscosity and $\kappa$ as the thermal diffusivity, all assumed to be independent of temperature. The Rayleigh and Prandtl numbers are then respectively defined as
\begin{subeqnarray} \label{eqn:RaAndPr}
	\gdef\thesubequation{\theequation \mbox{\textit{a}},\textit{b}}
	\Ray \equiv g\beta\Delta L^3/(\nu\kappa),\quad \Pran \equiv \nu/\kappa,
\end{subeqnarray}
\returnthesubequation
and the Nusselt and Reynolds numbers are respectively defined as
\begin{subeqnarray} \label{eqn:NuAndRe}
	\gdef\thesubequation{\theequation \mbox{\textit{a}},\textit{b}}
	\Nus \equiv JL/(\Delta \kappa),\quad \Rey \equiv UL/\nu,
\end{subeqnarray}
\returnthesubequation
where $J \equiv -\kappa (\romd \overline{\varTheta}/\romd z) + \overline{w'\varTheta'}$ the horizontal heat flux and $U$ is a characteristic velocity scale. Equations (\ref{eqn:GoverningEquations}$a$--$c$) have been numerically solved in \cite{Ng+Ooi+Lohse+Chung.2014} for no-slip and impermeable boundary conditions for the velocity field at the walls (in the plane $z=0$ and $z=L$) and periodic boundary conditions in the $x$- and $y$-directions. The resulting mean streamwise velocity component ($\overline{u}=u-u^\prime$) and mean temperature ($\overline{\varTheta} = \varTheta - \varTheta^\prime$) are statistically antisymmetric about the channel centreline, as illustrated in figure \ref{fig:VCSetup}($b$). Here, we denote time- and $xy$-plane-averaged quantities with an overbar, and the corresponding fluctuating part with a prime. In the channel-centre of VC, both $\romd \overline{u}/\romd z$ and $\romd \overline{\varTheta}/\romd z$ are finite and possesses the same sign. Note that in VC, $\overline{u}$ is a persistent non-zero mean quantity, which is different to RBC: for a sufficiently long time-average, it can be shown that the wall-parallel-averaged velocity components in RBC are statistically zero \citep[\eg][]{vanReeuwijk+Jonker+Hanjalic.2008}.

For the present study, we are interested in two numerical setups that are different from VC. The new setups are defined such that they allow us to directly test the $1/2$-power scaling relations described in (\ref{eqn:HalfPowerLaws}). From this line of reasoning, the associated governing equations of the new setups should be expected to obey the scaling arguments of \cite{Kraichnan.1962} and \cite{Grossmann+Lohse.2000}, and in the spirit of deriving (\ref{eqn:HalfPowerLaws}). In short, the key idea here is to design numerical setups that only solve the fluctuating components of VC, which conveniently emulates the turbulent bulk-dominated conditions expected in the asymptotic ultimate regime of thermal convection at very high $\Ray$ \citep{Grossmann+Lohse.2000}. To this end, we describe two setups for VC, \ie\, HVC and HVCws, which are inspired by the so-called homogeneous configurations for RBC of \cite{Lohse+Toschi.2003} and \cite{Calzavarini+others.2005,Calzavarini+others.2006}. Different to homogeneous RBC, the HVC and HVCws setups described in the following sections are subjected to a mean horizontal temperature (or buoyancy) gradient, which is orthogonal to gravity. 

\subsection{Homogeneous vertical natural convection with shear (HVC)}
\label{subsec:HVCshear}
For HVC, we assume that the flow is decomposed into constant mean 
gradients and fluctuations. These assumptions are notionally similar to the flow 
conditions in the channel-centre of VC, as illustrated in figures 
\ref{fig:VCSetup}($b$) and \ref{fig:VCSetup}($c$). To describe the numerical 
approach, we also make use of the equation of state for gases, 
$\beta(\varTheta-\varTheta_0) \approx -(1/\rho_0)(\rho-\rho_0)$ and introduce $b$ 
the buoyancy variable. Therefore, following \cite{Chung+Matheou.2012}, we write
\begin{subeqnarray} \label{eqn:MeanEqnsWithShear}
	-(g/\rho_0) (\rho-\rho_0) &=& N^2 x_3 + b^\prime, \\
	u_i &=& S \delta_{i1} x_3 + u_i^\prime, \\
	p + \rho_0 g x_1 &=& p^\prime
\end{subeqnarray}
\returnthesubequation
where $N^2 \equiv \romd\overline{b}/\romd z = g\beta \romd 
\overline{\varTheta}/\romd z$ the constant mean buoyancy gradient, $S \equiv \romd 
\overline{u}/\romd z$ the (temporally) uniform mean shear and $u_i^\prime$, 
$b^\prime$ and $p^\prime$ are the fluctuations of velocity, buoyancy and pressure, 
respectively. Substituting (\ref{eqn:MeanEqnsWithShear}) into 
(\ref{eqn:GoverningEquations}), we obtain
\begin{subeqnarray} \label{eqn:GovEqnWithShear}
	\dfrac{\partial u_j^\prime}{\partial x_j} &=& 0, \\
	\dfrac{\partial b^\prime}{\partial t} + N^2 u_3^\prime + 
	\dfrac{\partial u_j^\prime b^\prime}{\partial x_j} 
	+ S\delta_{j1}x_3 \dfrac{\partial b^\prime}{\partial x_j}
	&=& \kappa \dfrac{\partial^2 b^\prime}{\partial x_j^2} \\
	\dfrac{\partial u_i^\prime}{\partial t}
	+ S\delta_{i1}u_3^\prime + \dfrac{\partial u_j^\prime u_i^\prime}{\partial x_j}
	+ S \delta_{j1}x_3 \dfrac{\partial u_i^\prime}{\partial x_j}
	&=& -\dfrac{1}{\rho_0}\dfrac{\partial p^\prime}{\partial x_i}
	+ \nu\dfrac{\partial^2 u_i^\prime}{\partial x_j^2} 
	+ (N^2x_3 + b^\prime) \delta_{i1}.
\end{subeqnarray}
\returnthesubequation
The methodology to solve 
(\ref{eqn:GovEqnWithShear}$a$--$c$) closely follows the approach by
\cite{Chung+Matheou.2012}, that is, a skewing coordinate $\xi_i (x_i,t) = x_i - S 
t\delta_{i1}x_3$ is introduced to transform the dependent variables 
$\{b',u_i',p'\}(x_i,t) = \{\widetilde{b}, \widetilde{u}_i, \widetilde{p} \} ( 
\xi_i(x_i,t),t)$ to convert (\ref{eqn:GovEqnWithShear}) to
\begin{subeqnarray} \label{eqn:GovEqnXiWithShear}
	\dfrac{\partial \widetilde{u}_j}{\partial \widetilde{\xi}_j} &=& 0, \\
	\dfrac{\partial \widetilde{b}}{\partial t} + N^2\widetilde{u}_3 + 
	\dfrac{\partial \widetilde{u}_j \widetilde{b}}{\partial \widetilde{\xi}_j}
	&=& \kappa \dfrac{\partial^2 \widetilde{b}}{\partial \widetilde{\xi}_j^2}, \\
	\dfrac{\partial \widetilde{u}_i}{\partial t}
	+ S\delta_{i1}\widetilde{u}_3 + \dfrac{\partial \widetilde{u}_j 
	\widetilde{u}_i}{\partial \widetilde{\xi}_j}
	&=& -\dfrac{1}{\rho_0}\dfrac{\partial \widetilde{p}}{\partial \widetilde{\xi}_i}
	+ \nu\dfrac{\partial^2 \widetilde{u}_i}{\partial \widetilde{\xi}_j^2}
	+ (N^2 \xi_3 + \widetilde{b}) \delta_{i1},
\end{subeqnarray}
\returnthesubequation
where $\partial/\partial \widetilde{\xi}_i \equiv \partial/\partial \xi_i - St 
\delta_{i3} \partial/\partial \xi_1$. 

The transformation from (\ref{eqn:GovEqnWithShear}) to (\ref{eqn:GovEqnXiWithShear}) allows us to numerically solve (\ref{eqn:GovEqnXiWithShear}) in a triply periodic domain provided $N^2\xi_3$ the non-periodic term on the right-hand-side of (\ref{eqn:GovEqnXiWithShear}$c$), which acts on the streamwise momentum, can be neglected. This is satisfied if $\vert N^2 \xi_3\vert \ll \vert \widetilde{b}\vert$. Estimating $\vert \widetilde{b}\vert = O(b_{rms})$, the inequality then holds true for scales in the $z$-direction that are smaller than $b_{rms}/N^2 (= \varTheta_{rms} (\romd \overline{\varTheta}/\romd z)^{-1})$. There is no straightforward method to determine beforehand if $b_{rms}/N^2$ is larger or smaller than $L_z$ without performing the homogeneous simulations. As a start, we omit the non-periodic term from our DNS, noting that this is a necessary numerical approximation. On the other hand, if $b_{rms}/N^2 < L_z$, the solutions based on the DNS without the non-periodic term are still meaningful, provided we focus only on the dynamics of the scales that are $\lesssim b_{rms}/N^2$. Indeed, we will show in \S\,\ref{subsec:Assumptions} that $b_{rms}/N^2 < L_z$ for the homogeneous cases, which we then further enforce in our calculations of the Nusselt and Reynolds number in \S\,\ref{subsec:ScalingOfNuAndRe}, where we apply a spectral filter to the Nusselt and Reynolds number using a cut-off length of the order of $b_{rms}/N^2$, smaller than $L_z$.

The solutions to (\ref{eqn:GovEqnXiWithShear}) without the non-periodic term describe the evolution of the fluctuating quantities of VC under the influence of a prescribed mean buoyancy gradient and mean shear. We acknowledge that the aforementioned assumptions are merely simplifications since both the mean shear $S$ and the buoyancy gradient $N^2$ are in principle the responding parameters of the bulk flow of VC; the boundary layers that form at the walls determine $S$ and $N^2$. Thus, an explicit relation between $S$ and $N^2$ is presently unknown, at least to our knowledge. Similarly, in the case of HVC, both $S$ and $N^2$ are not known \textit{a priori} but must be prescribed. (This is detailed in \S\,\ref{sec:ComputParam}). We emphasize that the formulation for HVC above is not an attempt to simulate the bulk flow of VC --- it is instead an idealised numerical model that is designed to test the $1/2$-power-law scalings of (\ref{eqn:HalfPowerLaws}).

\subsection{Homogeneous vertical natural convection without shear 
(HVCws)} \label{subsec:HVCnoshear}
In addition to HVC, HVCws is an alternative numerical setup that can test the $1/2$-power-law scalings by assuming that the mean shear component $S$ in (\ref{eqn:MeanEqnsWithShear}$b$) is zero, \ie\,$u_i = u_i^\prime$. It follows that the homogeneous simulations without shear can be conducted in the same triply periodic domain as HVC, but with the second term on the left-hand-side of (\ref{eqn:GovEqnXiWithShear}$c$), which is $S\delta_{i1}\widetilde{u}_3$, set to zero.

This assumption of the zero-mean shear is inspired by similar homogeneous studies for RBC \citep[\eg][]{Lohse+Toschi.2003,Calzavarini+others.2005}. For VC, the zero-mean-shear assumption is evidently fictitious since in reality, a mean flow is present. However, we make this assumption for the sake of convenience since we will show later in \S\,\ref{subsec:RaScalingHVCNoShear} that, with $S=0$, the $1/2$-power-law scaling arguments appear to hold more naturally. Note that for past studies on homogeneous RBC \citep[\eg][]{Lohse+Toschi.2003,Calzavarini+others.2005}, the zero-mean-shear assumption inherently holds because the mean velocity components for RBC are zero. The zero-mean-shear assumption in RBC also implies that, in principle, the homogeneous RBC setup would be directly comparable to the HVCws setup, in contrast to the more phenomenologically accurate HVC setup.

\section{The relationship between the homogeneous setups and the 
1/2-power-law asymptotic ultimate scaling} \label{sec:UltimateScalingInHVC}
\subsection{Definitions of the dimensionless numbers for the bulk}

Based on the setup for HVC and HVCws, we will now attempt to establish \textit{a priori} the expected power-law scaling for $\Nus$ and $\Rey$ in terms of $\Ray$ and $\Pran$. Specifically, we are interested to determine whether the $1/2$-power-law scalings described in (\ref{eqn:HalfPowerLaws}) could also be expected for the homogeneous cases. Our approach follows the same scaling arguments as described in \cite{Grossmann+Lohse.2000} for the asymptotic ultimate regime, which is referred in their work as the bulk-dominated regime for low-$\Pran$ thermal convection (regime $IV_l$), \ie\,when $\Pran \leqslant 1$.

Before proceeding further, we first need to redefine the Rayleigh, Nusselt and Reynolds numbers for the homogeneous setups, \ie\,(\ref{eqn:RaAndPr}$a$) and (\ref{eqn:NuAndRe}), since the temperature scale $\Delta$ and the characteristic velocity scale $U$ are undefined for HVC and HVCws. The temperature scale $\Delta$ in the Nusselt and Rayleigh numbers refers to the imposed temperature difference, and so we adopt $\Delta \equiv \Delta_b \equiv -(\romd \overline{\varTheta}/\romd z)L$. The velocity scale $U$ in the Reynolds number measures the system response, which are the velocity fluctuations and so we can define $U \equiv u_{rms}$, where $u_{rms}$ is the time- and volume-averaged root-mean-square of the streamwise velocity fluctuations. Therefore, we recast $\Ray$, $\Nus$ and $\Rey$ for the homogeneous setups as
\begin{subeqnarray}\label{eqn:BulkRaReNu}
	\gdef\thesubequation{\theequation \mbox{$a$,$b$,$c$}}
	\Ray_b \equiv g\beta \Delta_b L^3/(\nu\kappa),\quad
	\Nus_b \equiv JL/(\Delta_b \kappa),\quad \Rey_b \equiv u_{rms}L/\nu.
\end{subeqnarray}
\returnthesubequation
To distinguish (\ref{eqn:BulkRaReNu}) from the definitions for VC, we adopt the subscript $b$ to refer to the bulk-related quantities in the homogeneous flow. The lengthscale parameter $L$ is presently undefined, but for similar homogeneous studies of shear turbulence, \cite{Sekimoto+Dong+Jimenez.2016} have shown that homogeneous flows are always `minimal' and constrained by the shortest domain length. As such, we will later employ this definition for $L$ in \S\,\ref{sec:ComputParam} for computing (\ref{eqn:BulkRaReNu}), but in the scaling arguments to follow, a different choice of $L$ simply affects the prefactors of the scaling arguments and not the exponent of the power law. Therefore, for the purposes of this section, the choice of $L$ is immaterial.

\subsection{$\Ray$-scaling in HVC} \label{subsec:RaScalingHVC}
Starting with HVC, we consider the time- and volume-averaged kinetic and thermal dissipation rates which are obtained by manipulating (\ref{eqn:GovEqnWithShear}) without the non-periodic term on the right-hand-side of (\ref{eqn:GovEqnWithShear}$c$),
\begin{subeqnarray}\label{eqn:DissipHVC}
	\gdef\thesubequation{\theequation \mbox{$a$,$b$}}
	\langle\varepsilon_{u'} \rangle = \beta g\langle u'\varTheta'\rangle - S 
	\langle u'w' \rangle, \qquad
	\langle\varepsilon_{\varTheta'} \rangle = (\Delta_b/L)\langle 
	w'\varTheta'\rangle.
\end{subeqnarray}
\returnthesubequation
The notation $\langle (\cdot) \rangle$ denotes time- and volume-averaging. Next, we write (\ref{eqn:DissipHVC}) explicitly in terms of (\ref{eqn:BulkRaReNu}). However, before doing so, we recognise that the two terms on the right-hand-side of (\ref{eqn:DissipHVC}$a$) are not know explicitly in terms of (\ref{eqn:BulkRaReNu}). Thus, we make two necessary assumptions which we verify later in \S\,\ref{subsec:Assumptions}: the first assumption is that $\langle u'\varTheta'\rangle \sim \langle w'\varTheta'\rangle$ and the second assumption is that $\langle u'w'\rangle \sim u_c^2 \equiv  \langle\varepsilon_{u'} \rangle/|S|$, where $u_c$ is the Corrsin velocity scale. 

A physical interpretation of the first assumption is warranted at this point: Because the driving heat flux is perpendicular to the gravity vector in our setup, we assume that a relatively greater and uniform mixing is present in HVC compared to homogeneous RBC. Thus, the HVC flow presumably generates vertical and horizontal small scales (in the direction of the driving heat flux) that are magnitude-wise comparable. A careful comparison between HVC and homogeneous RBC datasets at matched $\Ray_b$ is warranted to verify this relation. The first assumption is also felicitous and essential since the relation between the turbulent horizontal heat flux and the turbulent vertical heat flux is inherently unknown for VC, which is in contrast to RBC where both the turbulent driving and responding heat fluxes are parallel to gravity.

With the two assumptions, (\ref{eqn:DissipHVC}$a$) can be written as $\langle\varepsilon_{u'} \rangle \sim \beta g\langle w'\varTheta'\rangle - S (\langle\varepsilon_{u'} \rangle/|S|)$ and so we can explicitly write (\ref{eqn:DissipHVC}) as
\begin{subeqnarray}\label{eqn:DissipHVCStep1}
	\gdef\thesubequation{\theequation \mbox{\textit{a}},\textit{b}}
	\langle\varepsilon_{u'} \rangle \sim 
	\dfrac{\nu^3}{L^4}\Ray_b \Pran^{-2}(\Nus_b-1), \qquad
	\langle\varepsilon_{\varTheta'} \rangle = \kappa \dfrac{\Delta_b^2}{L^2} 
	(\Nus_b-1).
\end{subeqnarray}
\returnthesubequation

Next, we model the global-averaged dissipation rates on the left-hand-side of (\ref{eqn:DissipHVC}) following the dimensional arguments for the turbulence cascade in fully developed turbulence, where the dissipation rate of turbulent fluctuations scale with the energy of the largest eddies of the order of $u_{rms}^2$ over a time scale $L/u_{rms}$ \citep[][Chapter 6]{Pope2000turbulent}. By analogy, the dissipation rate of thermal variance scales with the largest eddies with variance $\varTheta_{rms}^2$ over a time scale $L/u_{rms}$. Thus,
\begin{subeqnarray} \label{eqn:DissipHVCStep2}
	\gdef\thesubequation{\theequation \mbox{\textit{a}},\textit{b}}
	\langle\varepsilon_{u'} \rangle 
		\sim \dfrac{u_{rms}^3}{L} 
		= \dfrac{\nu^3}{L^4}\Rey_b^3, \qquad
	\langle\varepsilon_{\varTheta'} \rangle 
		\sim \dfrac{u_{rms}\varTheta_{rms}^2}{L} 
		= \kappa\dfrac{\varTheta_{rms}^2}{L^2}\Pran \Rey_b.
\end{subeqnarray}
\returnthesubequation
\citep[cf.\,][]{Grossmann+Lohse.2000}. We can now match (\ref{eqn:DissipHVCStep1}$a$) with (\ref{eqn:DissipHVCStep2}$a$) and (\ref{eqn:DissipHVCStep1}$b$) with (\ref{eqn:DissipHVCStep2}$b$) and eliminate common terms to obtain,
\begin{subeqnarray} \label{eqn:HVCScaling1}
	\gdef\thesubequation{\theequation \mbox{\textit{a}},\textit{b}}
	\Ray_b \Pran^{-2}(\Nus_b-1) \sim \Rey_b^3, \qquad 
	\Delta_b^2(\Nus_b-1) \sim \varTheta_{rms}^2 \Pran \Rey_b.
\end{subeqnarray}
\returnthesubequation
Equation (\ref{eqn:HVCScaling1}$b$) can be simplified if we assume that $\Delta_b \sim \varTheta_{rms}$ and thus, we can manipulate (\ref{eqn:HVCScaling1}) to obtain
\begin{subeqnarray} \label{eqn:HVCScaling2}
	\gdef\thesubequation{\theequation \mbox{\textit{a}},\textit{b}}
	\Nus_b \sim \Ray_b^{1/2}\Pran^{1/2}, \qquad 
	\Rey_b \sim \Ray_b^{1/2}\Pran^{-1/2},
\end{subeqnarray}
\returnthesubequation
which is the same as the $1/2$-power-law expressions derived in equations (2.19) and (2.20) of \cite{Grossmann+Lohse.2000}, similar to the asymptotic Kraichnan regime \citep{Kraichnan.1962}.

Alternatively, we can emphasize the role of the mean components on the turbulent dissipation rates \citep[as shown in][]{Ng+Ooi+Lohse+Chung.2014} by defining the energy of the kinetic and thermal eddies based on $u_b$ and $\Delta_b$, where $u_b \equiv -SL_z$. Therefore, instead of (\ref{eqn:DissipHVCStep2}), the global-averaged dissipation rates on the left-hand-side of (\ref{eqn:DissipHVC}) may be modelled as
\begin{subeqnarray} \label{eqn:DissipHVCStep3}
	\gdef\thesubequation{\theequation \mbox{\textit{a}},\textit{b}}
	\langle\varepsilon_{u'} \rangle 
	\sim \dfrac{u_{b}^3}{L} 
	= \dfrac{\nu^3}{L^4}\Rey_b^3 \left(\dfrac{u_b^3}{u_{rms}^3}\right),\quad
	\langle\varepsilon_{\varTheta'} \rangle 
	\sim \dfrac{u_b \Delta_b^2}{L} 
	= \kappa\dfrac{\Delta_b^2}{L^2} \Pran \Rey_b \left( \dfrac{u_b}{u_{rms}} \right).
\end{subeqnarray}
\returnthesubequation
Equation (\ref{eqn:DissipHVCStep3}) can be simplified if we assume $u_b \sim u_{rms}$. Thus, we can match (\ref{eqn:DissipHVCStep1}$a$) with (\ref{eqn:DissipHVCStep3}$a$) and (\ref{eqn:DissipHVCStep1}$b$) with (\ref{eqn:DissipHVCStep3}$b$), as before, and recover the same $1/2$-power-law scaling of (\ref{eqn:HVCScaling2}). Both assumptions $\Delta_b\sim\varTheta_{rms}$ and $u_b \sim u_{rms}$ are reasonable in the absence of walls, since the fluctuating quantities respond directly to the input quantities $\Delta_b$ and $u_b$, which are constant \citep{Calzavarini+others.2005}. When walls are present, a different treatment is necessary and would depend on the distance from the wall, see for example the mixing length model proposed in \cite{Shishkina+Others.2017}.

In summary, the governing equations for HVC appear to provide a natural $1/2$-power-law scaling in the spirit of the GL-theory formulation. However, we reiterate that the homogeneous setup is merely an idealisation which enables us to test the $1/2$-power-law scaling and does not explicitly model the flow at the channel-centre of VC. The scaling arguments above are consistent with the approach previously discussed by \cite{Lohse+Toschi.2003} for homogeneous RBC.

\subsection{$\Ray$-scaling in HVCws} \label{subsec:RaScalingHVCNoShear}
When the shear is absent in the homogeneous setup, the scaling arguments are relatively more straightforward because $S = 0$. That is, by manipulating (\ref{eqn:GovEqnWithShear}) without the terms containing $S$ and the non-periodic term, we obtain the global-averaged kinetic and thermal dissipation rates
\begin{subeqnarray}\label{eqn:DissipHVCNoShear}
	\gdef\thesubequation{\theequation \mbox{\textit{a}},\textit{b}}
	\langle\varepsilon_{u'} \rangle = \beta g\langle u'\varTheta'\rangle, \quad
	\langle\varepsilon_{\varTheta'} \rangle = (\Delta_b/L)\langle 
	w'\varTheta'\rangle.
\end{subeqnarray}
\returnthesubequation
We can now repeat the only assumption that $\langle u'\varTheta'\rangle \sim \langle w'\varTheta'\rangle$ to rewrite (\ref{eqn:DissipHVCNoShear}) explicitly as \begin{subeqnarray}\label{eqn:DissipHVCNoShearStep1}
	\gdef\thesubequation{\theequation \mbox{\textit{a}},\textit{b}}
	\langle\varepsilon_{u'} \rangle \sim \dfrac{\nu^3}{L^4}\Ray_b 
	\Pran^{-2}(\Nus_b-1),\quad \langle\varepsilon_{\varTheta'} \rangle = \kappa 
	\dfrac{\Delta_b^2}{L^2}(\Nus_b-1),
\end{subeqnarray}
\returnthesubequation
which is the same as (\ref{eqn:DissipHVCStep1}) for HVC. 

Next, since only the fluctuations are relevant for HVCws, the global-averaged dissipation rates on the left-hand-side of (\ref{eqn:DissipHVCNoShear}) scale according to (\ref{eqn:DissipHVCStep2}). Lastly, by matching (\ref{eqn:DissipHVCNoShearStep1}) and (\ref{eqn:DissipHVCStep2}), and manipulating the dimensionless terms, we again obtain 
\begin{subeqnarray} \label{eqn:HVCNoShearScaling1}
	\gdef\thesubequation{\theequation \mbox{\textit{a}},\textit{b}}
	\Nus_b \sim \Ray_b^{1/2}\Pran^{1/2}, \qquad 
	\Rey_b \sim \Ray_b^{1/2}\Pran^{-1/2},
\end{subeqnarray}
\returnthesubequation
which is the same as (\ref{eqn:HVCScaling2}).

From the scaling arguments above, we conclude that both HVC and HVCws are expected to exhibit $1/2$-power-law scaling exponents, provided the assumptions that $\langle u'\varTheta'\rangle \sim \langle w'\varTheta'\rangle$, $\Delta_b \sim \varTheta_{rms}$ and $u_b \sim u_{rms}$ hold.

\section{Computational parameters} \label{sec:ComputParam}

\begin{table}
	\def\drawline#1#2{\raise 2.5pt\vbox{\hrule width #1pt height #2pt}}
	\def\spacce#1{\hskip #1pt}
	\begin{center}
		\def~{\hphantom{0}}
		\begin{tabular}{cccccccccccccc}
			& \multicolumn{6}{c}{HVC} & & \multicolumn{6}{c}{HVC without shear} \\ 
			& \multicolumn{6}{c}{{\color{black}\hbox{\drawline{120}{0.5}}}} & 
			{\color{white}\hbox{\drawline{5}{0.5}}}& 
			\multicolumn{6}{c}{{\color{black}\hbox{\drawline{120}{0.5}}}} \\ [3pt]
			$\Ray_b$ 		& $N_x$	& $N_y$	& $N_z$ & $k_{\text{max}}\eta$ & 
			$N^2/S^2$		&	$T^\ast$ & & $N_x$	& $N_y$	& $N_z$ & 
			$k_{\text{max}}\eta$ & 
			$N^2/S^2$		&	$T^\ast$ \\ [3pt]
			$5\times 10^4$	&  ~256 &	~64	&	128	&  	4.06 &  0.256 &  751 & &  
							    384 &	~96	&	192	&  	5.40 &	inf &  397 \\
			$1\times 10^5$	&  ~256 &	~64	&	128	&  	2.50 &	0.239 &  213 & &  
							    256 &	~64	&	128	&  	2.51 &	inf &  217 \\
			$4\times 10^5$	&  ~512 &	128	&	256	&  	3.10 &	0.175 &  218 & &  
							    512 &	128	&	256	&  	3.39 &	inf &  240 \\
			$1\times 10^6$	&  ~512 &  	128	&	256	&	2.24 & 	0.141 &  206 & &  
							    512 &	128	&	256	&  	2.45 &	inf &  867 \\
			$4\times 10^6$	&  ~768 & 	192	&	384	&	2.06 &  0.115 &  201 & &  
							    512 &	128	&	256	&  	1.48 &	inf &  207 \\
			$1\times 10^7$	&  1024 &	256	&	512	& 	1.64 &  0.117 &  204 & &  
							    768 &	192	&	384	&  	1.34 &	inf &  210 \\
			[3pt]
		\end{tabular}
		\caption{Simulation parameters of the present DNS cases for HVC and HVC 
			without shear. For all cases, $L_x = 4L_y = 2L_z$. Here, 		
			$k_{\text{max}}$ the maximum dealiased wavenumber magnitude and 
			$T^\ast \equiv T_{\textit{samp}} U_{\Delta}/L_y$, where $T_{samp}$ is 
			the sampling interval.}
		\label{tab:SimParam}
	\end{center}
\end{table}

We now proceed to describe the numerical simulations of the homogeneous setups. The simulations are performed in a triply periodic box with a height that is twice its horizontal width, but with its depth that is half of its horizontal width: $L_x = 4L_y = 2L_z$ (see figure \ref{fig:VCSetup}$c$ and $d$). The longer streamwise length is in anticipation of the streamwise-elongated structures in the presence of shear \citep{Chung+Matheou.2012}, \ie\,for the case of HVC, whereas the shorter spanwise length used in our simulations fulfils the limits proposed for stationary shear flows \citep[\eg][]{Sekimoto+Dong+Jimenez.2016}. A qualitative assessment on the sensitivity of the domain sizes is provided in Appendix \ref{sec:AppendixA}. Both the longer streamwise box height and shorter spanwise depth are necessary for the homogeneous setup in the presence of shear, which makes our domain different to previous studies on homogeneous thermal convection, such as for homogeneous RBC which were conducted in a box with equal height, width and depth \citep[\eg][]{Lohse+Toschi.2003,Calzavarini+others.2005,Calzavarini+others.2006}. For consistency and to assist comparison, we employ the box dimensions for HVC to the cases of HVCws. Since $L_y$ is the shortest box dimension for our setup, the largest structures of the flow are determined by this domain length (\cf\,Appendix \ref{sec:AppendixA}) and so, we adopt $L\equiv L_y$ in our definition of $\Ray_b$ in (\ref{eqn:RaAndPr}$a$) and also for the simulation parameters described below. The spanwise-domain-based definition of $\Ray_b$ for the homogeneous setups also allows for a meaningful comparison with the bulk of VC and the associated channel-width-based definition of the Rayleigh number, where the bulk flow is determined by the distant boundary layers. All of our simulations employ equal grid spacings in all three directions. 

\begin{figure}
	\centering
	\centerline{\includegraphics{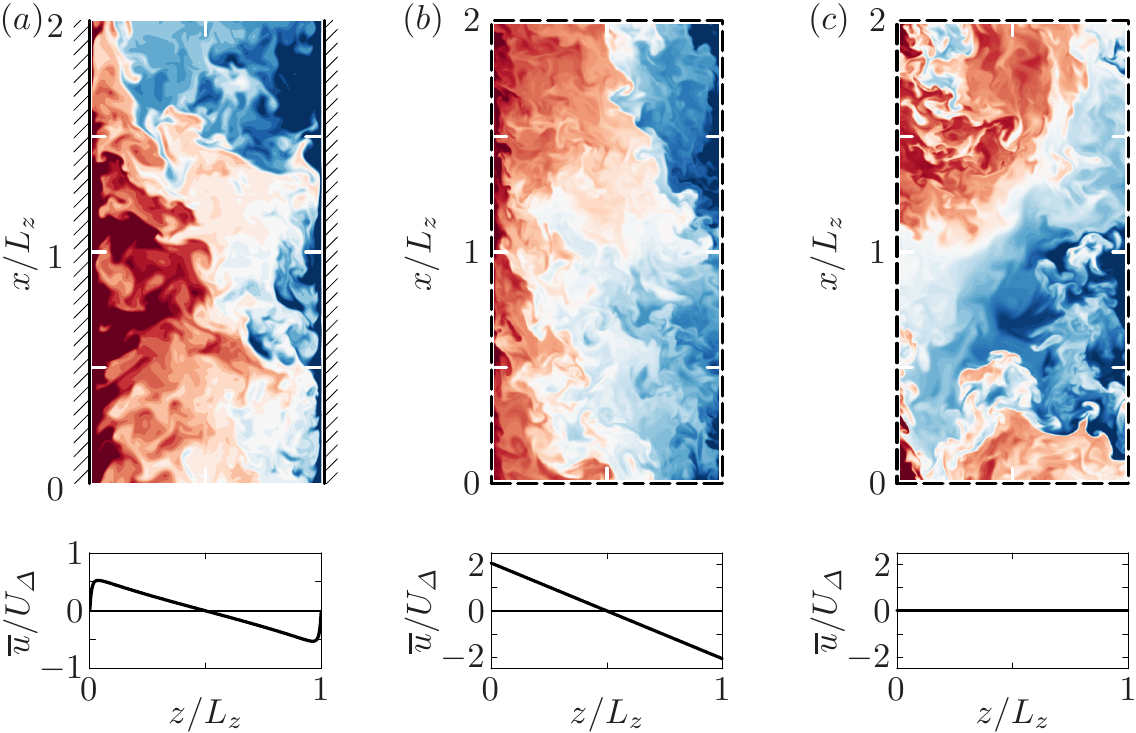}}
	\caption{\label{fig:TempVisuals}Streamwise-wall-normal visualisations of 
	$\varTheta$ the	instantaneous temperature field at matched Rayleigh number 
	($\Ray_b \approx 10^7$) for ($a$) VC, ($b$) HVC and ($c$) HVCws, 
	highlighting the difference between homogeneous setups and VC. Temperature 
	is decreasing from red to blue and the three plots share the same 
	colour map. The tilted structures in ($b$) are reminiscent 
	of the tilted structures in ($a$) at $z \approx 0.5L_z$, but the structures in 
	($c$) bear little resemblance to the structures in ($a$) and ($b$). The 
	respective mean velocity profiles are shown in the subplots beneath the 
	visualisations. As in figure \ref{fig:VCSetup}, the hatched boundaries in ($a$) 
	represent walls and the dashed boundaries in ($b$) and ($c$) represent periodic 
	boundary conditions. Note that in the visualisation in ($a$) only a quarter 
	section of VC is reproduced from the DNS of \cite{Ng+Ooi+Lohse+Chung.2014}.}
\end{figure}

Table \ref{tab:SimParam} summarises the relevant simulation parameters used in this study. The simulation adopts a Fourier pseudospectral method (with the typical 2/3-rule for dealiasing in wavenumber space) that is stepped in time using the low-storage third-order Runge--Kutta scheme \citep[cf.][]{Chung+Matheou.2012} at the timestepping interval $\Delta_t=CFL~\text{max}_i(\Delta_i/u_i')$, where we have set $CFL=0.9$ and $\Delta_i$ is the dealiased grid spacing in each direction. The data from the simulations are sampled for at least approximately 200 turnover times based on the free-fall period $t_f \equiv L_y/U_{\Delta}$, where $U_{\Delta} \equiv \sqrt{g\beta\Delta_b L_y}$, after discarding at least 99 turnover times. When compared to the diffusive timescale $t_\kappa \equiv L_y^2/\kappa$, it can be shown that $t_\kappa = (\Ray_b\Pran)^{1/2}t_f$, reflecting that the free-fall timescale is much faster than the diffusive timescale by a factor of $(\Ray_b\Pran)^{1/2}$. In the present study, the simulations are resolved up to at least $k_{\text{max}}\eta \approx 1.34$, where $k_{\text{max}}$ is the maximum dealiased wavenumber magnitude, $\eta \equiv (\nu^3/\langle\varepsilon_{u'}\rangle )^{1/4}$ is the Kolmogorov scale and $\langle\varepsilon_{u'}\rangle \equiv \nu \langle (\partial u'_i/\partial x_j)^2 \rangle$. The values of 	$\langle\varepsilon_{u'}\rangle$ used in table \ref{tab:SimParam} and throughout this paper are calculated explicitly from the gradients of the velocity fluctuations in our DNS. For a comparison with the approximations for $\langle\varepsilon_{u'}\rangle$, \ie\,the right-hand-side of (\ref{eqn:DissipHVC}$a$) in the case of HVC and the right-hand-side of (\ref{eqn:DissipHVCNoShearStep1}$a$) in the case of HVCws, we have included a discussion in Appendix \ref{sec:AppendixB}. For HVCws, we selected a finer resolution for the lowest $\Ray_b$ case ($=5\times10^4$), such that it is comparable to the resolution used in past numerical studies of low-$\Ray_b$ for homogeneous RBC \citep[e.g.\,][]{Calzavarini+others.2006}. We also note that for HVCws, different choices of resolution will affect the maximum value $\Nus_b$. The counter-intuitive matter of grid sensitivity has been previously highlighted, for example by \cite{Calzavarini+others.2006}, and also found and explained by \cite{Schmidt+others.2012} for axially homogeneous RBC. Similar to \cite{Schmidt+others.2012}, we find that the value of $\Nus_b$ is lower when a coarser resolution is used. For consistency with previous studies, here we only report the results from higher-resolution simulations.

For the HVC cases, 2 additional parameters need to be determined, namely $N^2$ and $S$. The selection of the values of $N^2$ and $S$ for the HVC cases are not intuitive since there are no boundary layers to set the values, thus, an explicit relation between $N^2$ and $S$ as a function of $\Ray_b$ are unknown. For the lack of a better justification, we define $N^2$ and $S$ for HVC based on the channel-centre gradients for VC from the DNS dataset of \cite{Ng+Ooi+Lohse+Chung.2014}. That is, we define $N^2 \equiv g\beta (\romd \overline{\varTheta}/\romd z) \rvert_c$ and $S \equiv (\romd \overline{u}/\romd z) \rvert_c$, where $(\romd \overline{\varTheta}/\romd z) \rvert_c = -\Delta_b/L_z$ and the subscript $c$ denotes quantities at the channel-centre of VC. By doing so, we assume that the idealised HVC cases are driven by both $N^2$ and $S$, which is different to VC, where both $N^2$ and $S$ are the system responses. For all of our simulations, we hold both $N^2$ and $S$ constant with time.

To illustrate the differences between the wall-bounded and homogeneous cases, we visualise the instantaneous temperature fields for the homogeneous setups and VC at matched $\Ray_b$ in figure \ref{fig:TempVisuals}. The values of $N^2$ and $S$ for the HVC case in figure \ref{fig:TempVisuals}($b$) match those for VC in figure \ref{fig:TempVisuals}($a$). There are subtle differences in the structures of both flows. For example, both fields share a similar tilting of structures in the mid-plane region ($z/L_z \approx 0.5$), but for the VC flow, the tilting of the structures changes closer to the left- and right-edges of the figure, due to the impermeable wall boundary conditions, that is, boundary layers are present in figure \ref{fig:TempVisuals}($a$). In contrast, for HVCws (figure \ref{fig:TempVisuals}$c$), there are no obvious tilting of the structures due to the absence of shear. The differences in the homogeneous flows are to be expected since both setups are idealisations and are not models for the bulk flow of VC. In the next section, we validate our selection of the mean gradients by comparing the dynamical scales in the bulk of VC and in the homogeneous cases.

\section{Comparison of statistics between wall-bounded and homogeneous VC} 
\label{sec:CompareStatistics}

In this section, we rationalize our choices of $N^2$ and $S$ for simulating the homogeneous cases by comparing with the statistics in the bulk of VC. Strictly speaking and as highlighted earlier, the homogeneous flow does not replicate the flow in the bulk of VC and our interest is to employ the homogeneous setup as an alternative to investigate the $1/2$-power asymptotic ultimate regime scaling. However, the dynamical quantities from the homogeneous cases should be reasonably comparable to the quantities in bulk of VC in order to justify comparison. In that respect, we define the characteristic length scales
\begin{subeqnarray}\label{eqn:LengthScales}
	\gdef\thesubequation{\theequation \mbox{\textit{a},\textit{b},\textit{c}}}
	l_c \equiv (\langle\varepsilon_{u'}\rangle/|S|^3)^{1/2},
	\quad \eta \equiv (\nu^3/\langle\varepsilon_{u'}\rangle)^{1/4},
	\quad l_o \equiv (\langle\varepsilon_{u'}\rangle/|N|^3)^{1/2},
\end{subeqnarray}
\returnthesubequation
which are the Corrsin, Kolmogorov and Ozmidov-like length scales, respectively, following studies on stratified shear turbulence \citep[\eg][]{Smyth+Moum.2000,Chung+Matheou.2012}. Note that we define $l_o$ as a length scale that is analogous to the Ozmidov scale. We recognise that the definition is not perfect since the effects of stratification is absent in VC owing to the direction of the temperature gradient, which acts in the horizontal direction and is orthogonal to the direction of the vertical shear. Nevertheless, we employ $l_o$ simply as a convenient buoyancy length scale. $l_o$ is also defined modulo-wise because the horizontal buoyancy gradient $g\beta(\romd \overline{\varTheta}/\romd z)\vert_c$ is negative for VC. The length scales defined in (\ref{eqn:LengthScales}) can also be written in the form of the dimensionless parameters that define scale separations for stratified turbulence:
\begin{subeqnarray}\label{eqn:SeparationScales}
	\gdef\thesubequation{\theequation \mbox{\textit{a}},\textit{b},\textit{c}}
	\Ric \equiv (N^2/S^2) = (l_c/l_o)^{4/3}, \quad \R \equiv (l_o/\eta)^{4/3}, \quad 
	\Rey_s \equiv (l_c/\eta)^{4/3},
\end{subeqnarray}
\returnthesubequation
\citep[\cf][]{Chung+Matheou.2012} which are the gradient Richardson number, buoyancy Reynolds number and shear Reynolds number, respectively.

\begin{figure}
	\centering
	\centerline{\includegraphics{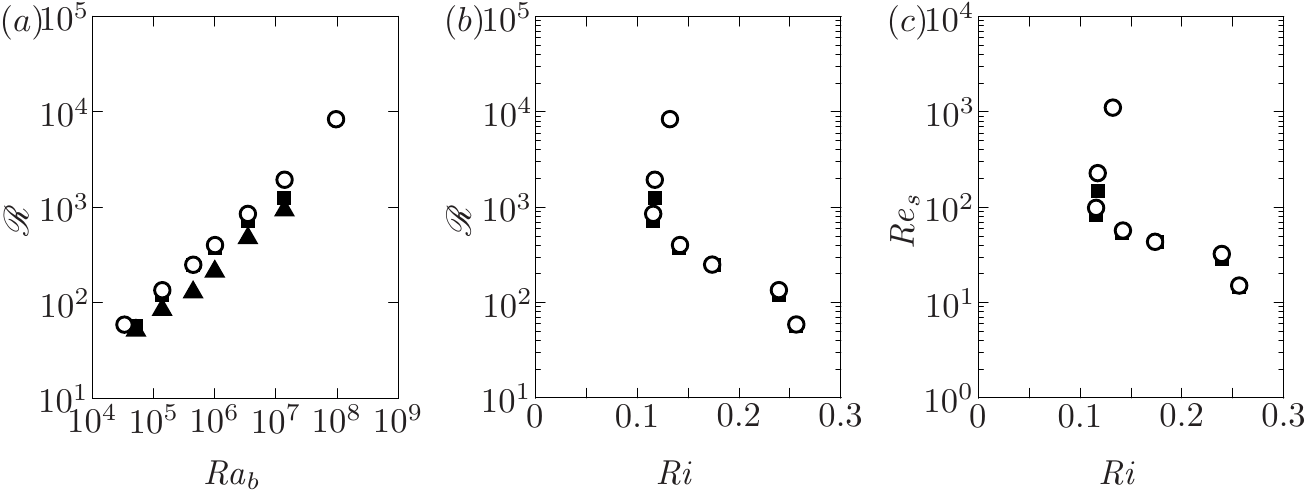}}
	\vspace*{1em}
	\caption{\label{fig:Lengthscale}Comparison of lengthscale separations in HVC ($\blacksquare$), HVCws ($\blacktriangle$) and bulk of VC ($\circ$) using ($a$) $\R$ versus $\Ray_b$, ($b$) $\R$ versus $\Ric$, and ($c$) $\Rey_s$ versus $\Ric$. The quantities $\R$, $\Ric$ and $\Rey_s$ are defined in (\ref{eqn:SeparationScales}). For HVC without shear, $\Ric = \text{inf}$, and 	are therefore not shown in ($b$) and ($c$).}
\end{figure}

Figure \ref{fig:Lengthscale}($a$) shows the trends of $\R$ versus $\Ray_b$ for the homogeneous cases and for the bulk of VC. Note that for VC, we define $L\equiv L_z$ the lengthscale for $\Ray_b$, since $L_z$ is the shortest domain length for this setup. From figure \ref{fig:Lengthscale}($a$), the trends of $\R$ for the homogeneous cases are comparable to the bulk of VC, suggesting that the choices of $N^2$ and $S$ based on the channel-centre gradients of VC are reasonable. Similarly, the trends of $\R$ versus $\Ric$ in figure \ref{fig:Lengthscale}($b$) and $\Rey_s$ versus $\Ric$ in figure \ref{fig:Lengthscale}($c$) are also approximately equal. The $\Ric$-values are plotted on the abscissae since they are pre-determined input parameters for HVC, whereas both $\R$ and $\Rey_s$ on the ordinates are responses of HVC. In contrast, all three parameters are the responses of the bulk of VC. The agreement of the magnitudes of (\ref{eqn:LengthScales}) in figures \ref{fig:Lengthscale}($b$) and \ref{fig:Lengthscale}($c$) for HVC and bulk of VC indicates that the turbulent quantities in HVC respond in a scale-wise similar manner to the turbulent quantities in the bulk of VC.

\begin{figure}
	\def\drawline#1#2{\raise 2.5pt\vbox{\hrule width #1pt height #2pt}}
	\def\spacce#1{\hskip #1pt}
	\centering
	\centerline{\includegraphics{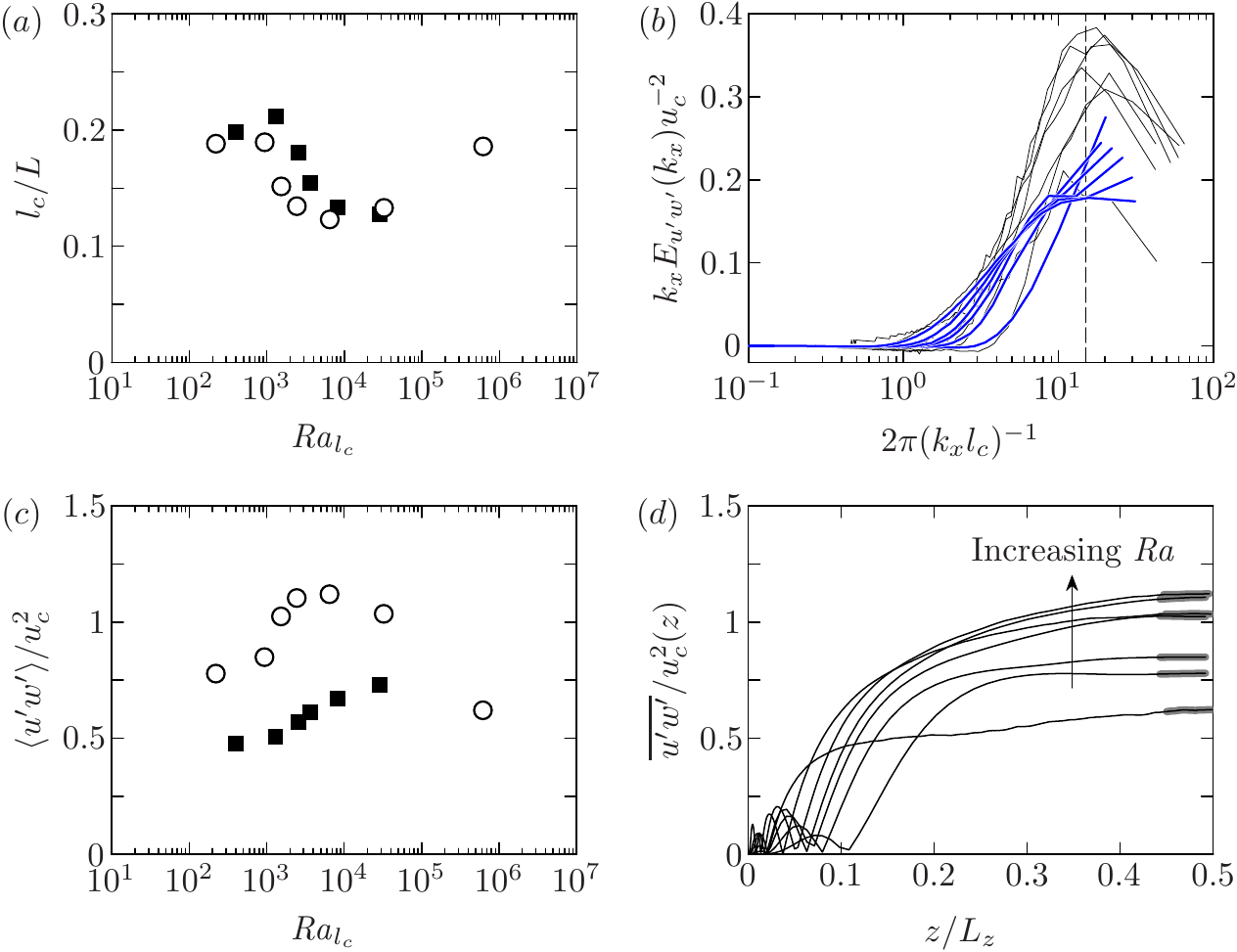}}
	\vspace*{1em}
	\caption{\label{fig:CorrsinScale}($a$) Comparison of the normalised Corrsin 	length $l_c/L$ versus $\Ray_{l_c}$ the Rayleigh number based on $l_c$, plotted 	for the bulk of VC ($\circ$) and HVC ($\blacksquare$). For VC, $L\equiv L_z$ and for HVC, $L \equiv L_y$. ($b$) Premultiplied one-dimensional spectra of 	$u'w'$ normalised by the Corrsin velocity $u_c$ for the bulk of VC 	({\color{black}\hbox{\drawline{8}{0.5}}}) and HVC 		({\color{blue}\hbox{\drawline{8}{0.5}}}). The vertical dashed line indicates the dominant wavelength of $\lambda_{x,p} \equiv 2\pi k_x^{-1} \approx 15l_c$. 	($c$) Ratio of the Reynolds shear stress to $u_c^2$ for the bulk of VC and HVC, where the symbols used are the same as in ($a$). ($d$) Similar to ($c$) for VC but plotted as a function of wall-normal location $z/L_z$. The average of the grey regions in ($d$) correspond to the values plotted for VC in ($c$). Note that although the highest $\Ray$ profile deviates from the overall trend the ratio remain roughly $\sim O(1)$.}
\end{figure}
In figure \ref{fig:CorrsinScale}($a$), we find that the magnitudes of the normalised Corrsin lengthscales for HVC and the bulk of VC are in good agreement. To facilitate comparison, we define the Rayleigh number for VC based on $l_c$. Noting also that $l_c$ is normalised by $L_y$ for HVC, and by $L_z$ for VC, the well-agreeing trends in figure \ref{fig:CorrsinScale}($a$) further suggests that $L_y$ can be considered as an appropriate characteristic lengthscale for the homogeneous cases, which is consistent with the limiting role of the shorter $L_y$ dimension discussed in \cite{Sekimoto+Dong+Jimenez.2016}. In figure \ref{fig:CorrsinScale}($b$), we compare the one-dimensional pre-multiplied spectra of $u'w'$ the Reynolds shear stress in the streamwise direction, normalised with $u_c$ the Corrsin velocity scale. Although the magnitude of the spectra for HVC is reduced owing to the restricted computational domain, both trends from HVC and VC roughly agree when scaled using the Corrsin units. The dominant streamwise wavelength corresponding to the peak in the pre-multiplied spectra is $\lambda_{x,p} \equiv 2\pi k_x^{-1} \approx 15l_c$ for both HVC and VC for the present $\Ray_b$-range. A more straightforward comparison between the Reynolds shear stress and Corrsin velocity scale is shown in figure \ref{fig:CorrsinScale}($c$), where we plot the values of $\langle u'w'\rangle/u_c^2$ versus $\Ray_{l_c}$ for both HVC and bulk of VC. In the case of VC, the values are calculated as the average in the channel-centre between $0.45 \leqslant z/L_z \leqslant 0.55$ where the ratio of $u'w'(z)/u_c^2(z)$ is approximately constant, as shown by the grey shaded lines in figure \ref{fig:CorrsinScale}($d$). The approximately constant trend of $u'w'(z)/u_c^2(z)$ close to the channel-centre of VC also implies that the Corrsin units are candidate scales for the bulk flow with shear. From both figures \ref{fig:CorrsinScale}($c$) and \ref{fig:CorrsinScale}($d$), we find that the fractions are both $\sim O(1)$, which suggest that $u_c$ is a reasonable measure of $u'w'$ for the HVC flow, \ie\,in support of our earlier assumption in \S\,\ref{subsec:RaScalingHVC} that $u^\prime w^\prime \sim u_c^2$.

The results from our analysis of the lengthscales are encouraging. Although the homogeneous setups are idealisations and do not represent the bulk flow of VC, the turbulent scales generated in the homogeneous cases are similar to the turbulent scales generated in the bulk of VC. Thus, keeping this insight in mind, we analyse and compare the scaling relations for VC and the homogeneous setups in the next section.

\section{Scaling relations in VC and homogeneous VC} 
\label{sec:ScalingRelations}

In the following, we test several assumptions using the results from the homogeneous cases. This is followed by the scaling of the Nusselt and Reynolds numbers, where we find reasonable agreement between the effective power-laws and the expected $1/2$-power-laws derived in \S\,\ref{sec:UltimateScalingInHVC}. Inspired by this insight, we extend our analysis by testing the scaling of $\Nus$ and $\Rey$ using the turbulent quantities in the bulk of VC.

\subsection{Validation of assumptions} \label{subsec:Assumptions}
\begin{figure}
	\centering
	\centerline{\includegraphics{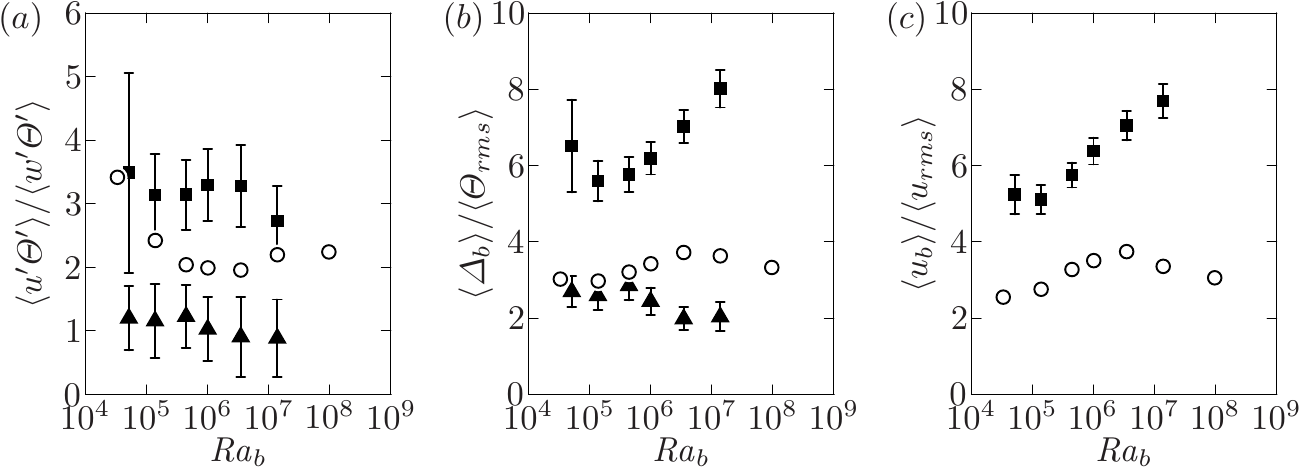}}
	\vspace*{1em}
	\caption{\label{fig:assumptions}Tests for the assumptions in 	\S\,\ref{sec:UltimateScalingInHVC}, using the trends of ($a$) $\langle 	u^\prime\varTheta^\prime\rangle/\langle w^\prime\varTheta^\prime\rangle$ versus $\Ray_b$, ($b$) $\langle\Delta_b\rangle/\langle\varTheta_{rms}\rangle$ versus 	$\Ray_b$, and ($c$) $\langle u_b\rangle/\langle u_{rms}\rangle$ versus $\Ray_b$. Cases shown are for HVC ($\blacksquare$), HVCws ($\blacktriangle$) and bulk of VC ($\circ$). The error bars represent half standard deviation from the mean of the fractions.}
\end{figure}
In this section we validate several assumptions employed in \S\,\ref{sec:UltimateScalingInHVC}, namely, the assumptions that $\langle u'\varTheta'\rangle \sim \langle w'\varTheta'\rangle$, $\Delta_b \sim \varTheta_{rms}$ and $u_b \sim u_{rms}$ for the homogeneous cases. The three trends are plotted in figure \ref{fig:assumptions} as fractions of their respective means with increasing $\Ray_b$. The solid square symbols represent the HVC case, solid triangle symbols for HVCws and open circles for VC. In figure \ref{fig:assumptions}($a$), the trends of $\langle u'\varTheta'\rangle/\langle w'\varTheta'\rangle$ are approximately constant for the $\Ray_b$ range for both homogeneous setups and for VC, which suggest that it is reasonable to assume that $\langle u'\varTheta'\rangle \sim \langle w'\varTheta'\rangle$. In contrast, both $\langle\Delta_b\rangle/\langle\varTheta_{rms}\rangle$ and $\langle u_b\rangle/\langle u_{rms}\rangle$ for HVC and HVCws in figures \ref{fig:assumptions}($b$) and \ref{fig:assumptions}($c$) exhibit weak $\Ray_b$-trends, suggesting that our scaling assumptions in \S\,3 may not be completely fulfilled. We reason that these weak $\Ray_b$-trends contaminate the effective power-law trends in the homogeneous cases resulting in a close-to-$1/2$ scaling exponent (shown later in \S\,\ref{subsec:ScalingOfNuAndRe}) as opposed to a definitive $1/2$-power-law scaling exponent.

From figure \ref{fig:assumptions}($b$), we can further analyse the trends of $\langle\Delta_b\rangle/\langle\varTheta_{rms}\rangle$ for both HVC and HVCws to substantiate our numerical approximation employed in our DNS (see \S\,\ref{subsec:HVCshear}). We not only require that $\Delta_b \sim \varTheta_{rms}$ to satisfy the scaling 	assumptions in \S\,\ref{sec:UltimateScalingInHVC}, but also that $\Delta_b/\varTheta_{rms}>1$ so that $b_{rms}/N^2 = (\Delta_b/\varTheta_{rms})^{-1}L_z < L_z$, which is modestly fulfilled by the trends shown in figure \ref{fig:assumptions}($b$). We acknowledge that whilst the values of the fractions are not significantly larger than 1, we contend with this limitation by restricting our analyses to only the energetic contributions from the lengthscales smaller than $L_z$, thereby also fulfilling the inequality. To that end, we apply a spectral filter in our calculations of the Nusselt and Reynolds numbers to discard the low-wavenumber (long-wavelength) contributions. In the subsequent sections, the filtered quantities are denoted with a superscript $^\ast$, where
\begin{equation}
(\cdot)^\ast = \int_{k_3^\ast}^{\infty} E_{(\cdot)}(k_3)~\romd 
k_3, \label{eqn:HiWavDefinition}
\end{equation}
$k_3$ the horizontal wavenumber and $2\pi (k_3^\ast)^{-1} = \lambda_3^\ast = 0.5L_z$.

\begin{figure}
	\centering
	\centerline{\includegraphics{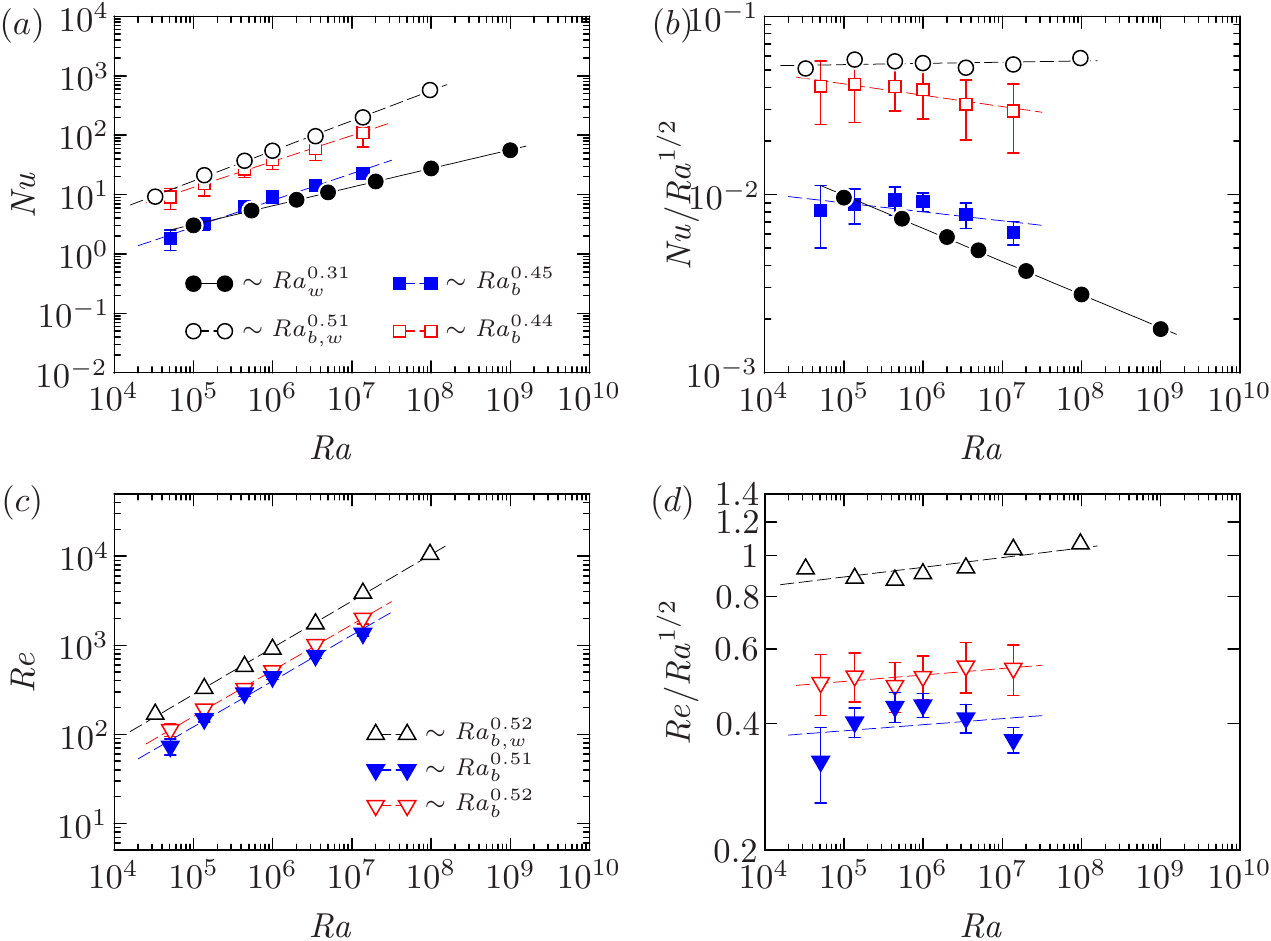}}
	\vspace*{1em}
	\caption{\label{fig:NuReRaScaling}($a$) Trends of Nusselt number versus Rayleigh number with slopes that are close to $1/2$ for the HVC case with shear ({\color{blue}$\fullsquare$}), without shear ({\color{red}$\opensquare$}) and bulk of VC ($\opencirc$). The slopes are steeper than the $\sim \Ray_w^{0.31}$ trend of the wall-gradient-based Nusselt number for VC ($\fullcirc$). ($c$) Same as in ($a$), but now reproduced for Reynolds number, again showing slopes that are close to $1/2$ for the HVC case with shear 	({\color{blue}$\fulltriangledown$}), without shear 	({\color{red}$\opentriangledown$}) and bulk of VC ($\opentriangle$). ($b$,$d$) Same trends as in ($a$) and ($c$) but compensated by the $1/2$-power-law for emphasis. Note that the Nusselt and Reynolds numbers for the HVC cases are based on the spectrally filtered small-scale quantities, as defined by equation (\ref{eqn:HiWavDefinition}). The error bars represent half standard deviation from the time-averaged Nusselt and Reynolds numbers.}
\end{figure}

\subsection{Scaling of $\Nus$ and $\Rey$} \label{subsec:ScalingOfNuAndRe}

We begin our scaling analyses by re-plotting the Nusselt number versus the Rayleigh number for VC, where $\Nus_w\equiv JL_z/(\Delta\kappa)$ and $\Ray_w \equiv g\beta\Delta L_z^3/(\nu\kappa)$, using the DNS data from \cite{Ng+Ooi+Lohse+Chung.2014}, where $(\cdot)_w$ denotes dimensionless quantities for VC. Figure \ref{fig:NuReRaScaling}($a$) shows the $\Nus_w$ versus $\Ray_w$ trend (black solid circles). When fitting a single power-law expression, $\Nus_w\sim\Ray_w^{\alpha_w}$, a least-squares fit results in $\alpha_w \approx 0.31$, obviously much smaller than the $1/2$-power-law exponent predicted in \S\,\ref{sec:UltimateScalingInHVC}. 

Now, we compare the effective exponent $\alpha_w$ to the effective exponents for the two homogeneous cases. We shall consider only the low-wavenumber-filtered Nusselt number, in accordance with our definition in (\ref{eqn:HiWavDefinition}), and the results of $\Nus_b^\ast$ versus $\Ray_b$ are plotted in figure \ref{fig:NuReRaScaling}($a$). With $\Nus_b^\ast\sim\Ray_b^{\alpha_{h}}$, the effective scaling exponent for HVC (solid blue squares) is $\alpha_{h} \approx 0.45$. The corresponding scaling for HVCws, using $\Nus_b^\ast\sim\Ray_b^{\alpha_{hns}}$ (open red squares) is $\alpha_{hns} \approx 0.44$. Both $\alpha_{h}$ and $\alpha_{hns}$ are steeper than $\alpha_w$, and are close to the $1/2$-power-law scaling. While not exactly equal to $1/2$, the close-to-$1/2$ effective power-law scalings in the homogeneous cases suggest that the homogeneous flow largely behaves in a manner similar to what is expected in the scaling arguments in \S\,\ref{sec:UltimateScalingInHVC}. The assumptions employed in \S\,\ref{sec:UltimateScalingInHVC} may also have a weak influence, which we discuss previously in \S\,\ref{subsec:Assumptions}. We accentuate the variations from the $1/2$-power-law by plotting the Nusselt number compensated by $\Ray^{-1/2}$ in figure \ref{fig:NuReRaScaling}($b$), where the non-horizontal trends from our homogeneous cases reflect the deviation from the $1/2$-power-law scaling. Our effective power-law scalings for the homogeneous cases are consistent with similar homogeneous studies on thermal convection but for different configurations, \ie homogeneous RBC \citep[e.g.\,][]{Lohse+Toschi.2003,Calzavarini+others.2005} and axially homogeneous RBC \citep[e.g.\,][]{Schmidt+others.2012}. Thus, it appears that the $1/2$-power-law dependency can be found in homogeneous VC where the thermal convection is quickly determined by the turbulent bulk.

Inspired by the bulk-scaling behaviour from the homogeneous cases, we now attempt to apply our understanding to VC. Unlike in RBC where the temperature gradient in the bulk is nominally zero \citep{Sun+Cheung+Xia.2008,Zhou+etal.2010} the temperature gradient in the bulk of VC is not. Therefore, we redefine the Rayleigh and Nusselt numbers for VC based on the quantities in the bulk of VC. Specifically, we define $\Delta_b = -L_z(\romd \overline{\varTheta}/\romd z)\rvert_c$ the bulk temperature scale in the channel-centre of VC. Additionally, since the horizontal heat flux is constant throughout the channel, the definition of $J$ for the bulk $\Nus$ is unchanged. Thus, the bulk quantities for VC can be defined by $\Nus_{b,w} \equiv \Nus_b$ and $\Ray_{b,w} \equiv \Ray_b$, and the trend is plotted in figure \ref{fig:NuReRaScaling}($a$) (black open circles). Interestingly, the least-squares fit to the power law $\Nus_{b,w} \sim \Ray_{b,w}^{\alpha_{b,w}}$ gives $\alpha_{b,w}\approx 0.51$ which is close to $1/2$ and of course much larger than the effective exponent $\alpha_{w} \approx 0.31$ discussed earlier. The horizontal trend is also illustrated in the compensated plot of figure \ref{fig:NuReRaScaling}($b$). It appears that the $1/2$-power scaling for bulk-dominated thermal convection may actually exist in all our previous simulations of VC even at low $\Ray$, if $\Delta_b$ is used instead of $\Delta$. However, further investigations, such as $\Pran$ regime studies, may be necessary to determine if this is indeed truly the case.

Moving on to the scaling of the Reynolds numbers, we plot $\Rey_b$ versus $\Ray_b$ for the homogeneous cases and for VC in figure \ref{fig:NuReRaScaling}($c$). For VC, we define the velocity scale as the root-mean-square of the streamwise velocity fluctuations at the channel-centre, \ie $U \equiv u_{rms,w} = (u^{\prime 2})^{1/2}\rvert_c$, and therefore $\Rey_{b,w} \equiv u_{rms,w}L/\nu$. In figure \ref{fig:NuReRaScaling}($d$), we accentuate the trends by plotting in the $1/2$-power compensated form. For the homogeneous cases, we again compute the low-wavenumber-filtered Reynolds number values, according to (\ref{eqn:HiWavDefinition}). For HVC, we obtain $\Rey_b^\ast \sim \Ray_b^{0.51}$ and for HVCws, we obtain $\Rey_b^\ast \sim \Ray_b^{0.52}$. The effective scaling exponents for the homogeneous cases are still in good agreement with the predictions in \S\,\ref{sec:UltimateScalingInHVC}. In addition, the exponents from the homogeneous cases are also close to the effective exponent from the bulk of VC, where we obtain $\Rey_{b,w} \sim \Ray_{b,w}^{0.52}$. The present results suggest that, even for VC, by selecting r.m.s.-based parameters to define the Reynolds number, we obtain an effective power-law that is consistent with the $1/2$-power-law scaling for bulk-dominated thermal convection. Indeed, our r.m.s.-based scaling results appear to be consistent not only with studies for RBC, \citep[\eg][]{vanReeuwijk+Jonker+Hanjalic.2008,Emran+Schumacher.2008} but also for the Rayleigh--Taylor flow, where light fluid is accelerated into heavy fluid \citep[\eg][]{Celani+Mazzino+Vozella.2006}.

As a final note, the $1/2$ effective power-law scaling results in figure \ref{fig:NuReRaScaling} for the homogeneous cases are not unexpected results; the homogeneous cases described by (\ref{eqn:GovEqnWithShear}) obey the $1/2$-power scaling arguments for the turbulent bulk-dominated regime as described by \cite{Grossmann+Lohse.2000} (see \S\,\ref{sec:UltimateScalingInHVC} above). As such, the homogeneous cases are merely alternative configurations to test the $1/2$-power-law scaling arguments but are not models for the bulk flow of VC.

\section{Exponential growth} \label{sec:ExponentialGrowth}
Homogeneous simulations of RBC have been reported to exhibit unstable and so-called `elevator modes' at low Rayleigh numbers, and are typically represented using exponentially growing values of the Nusselt number, followed by sudden break-downs \citep{Calzavarini+others.2005,Calzavarini+others.2006}. In this section, we examine the solutions of the homogeneous setup in order to determine whether such exponentially growing solutions also exist, and how the solutions compare to homogeneous RBC. To allow for a comparison between the homogeneous setup and homogeneous RBC, we restrict our analysis only to the case for HVCws (\cf\,\S\,\ref{subsec:HVCnoshear}).

\begin{figure}
	\centering
	\centerline{\includegraphics{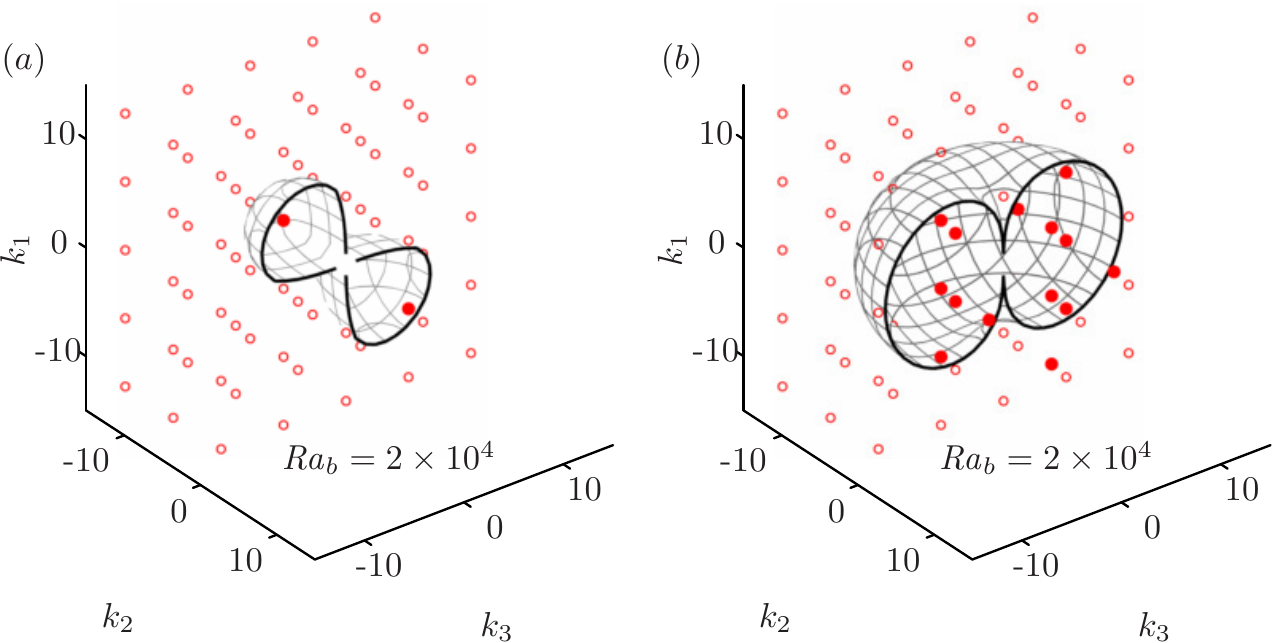}}
	\caption{\label{fig:3DModes}Isosurfaces of the marginally stable modes, which 	have $s_4=0$ for ($a$) HVCws and ($b$) homogeneous RBC at $\Ray_b=2\times10^4$ and $\Pran=0.709$. Inside these isosurfaces, the eigenvalues are positive (labelled ${\color{red}\bullet}$), corresponding to unstable modes. The number of unstable modes thus corresponds to the number of solid red markers enclosed within the isosurface: 2 for HVCws and 16 for homogeneous RBC (in the rotated view ($b$), only 14 are visible). Outside these isosurfaces, the eigenvalues are negative (labelled ${\color{red}\circ}$), corresponding to stable modes. Thanks to symmetry, the isosurface for $k_2>0$ do not need to be shown. The isosurfaces are oriented such that $k_1$ is associated with the physical space variable $x_1$, which is opposing gravity. The heat flux is in $x_3$-direction in ($a$) and in $x_1$-direction in ($b$). In ($a$), the isosurfaces shows a directional dependence in the ($k_2$, $k_3$) plane, in contrast to ($b$).}
\end{figure}
We begin our analysis by employing the method of small disturbances \citep[cf.\,\S\,2.5][]{Monin2007statisticalVol1} to the linearized equations of (\ref{eqn:GovEqnWithShear}), but without both the terms with shear and the non-periodic term in (\ref{eqn:GovEqnWithShear}$c$). That is, we set higher-order fluctuating terms to zero. When made dimensionless using the velocity scale $\kappa/L$, temperature scale $\Delta_b$ and the length scale $L$, which we represent by the notation $(\cdot)''$, the linear equations take the form
\begin{subeqnarray} \label{eqn:HVCStabEquations}
	\dfrac{\partial u''_i}{\partial x''_i} &=& 0, \\
	\dfrac{\partial \varTheta''_i}{\partial t''} &=&
	\dfrac{\partial^2 \varTheta''}{\partial x_j''^2} + u''_3 \\
	\dfrac{\partial u''_i}{\partial t''} &=& 
	-\dfrac{\partial p''}{\partial x''_i}
	+ \Pran \left(\dfrac{\partial^2 u''_i}{\partial x_j''^2} 
	+ \Ray_b\varTheta'' \delta_{i1}\right).
\end{subeqnarray}
\returnthesubequation
Next, we admit exponentially growing and time-dependent solutions of the form
\begin{subeqnarray}\label{eqn:expsol}
	\gdef\thesubequation{\theequation \mbox{\textit{a}},\textit{b}}
	\theta'' = \widehat{\theta}e^{st+\mathrm{i}\mathbf{k}\cdot \mathbf{x}} + 
	\mathrm{\,c.c.},
	\qquad
	u''_i		= \widehat{u}_ie^{st+\mathrm{i}\mathbf{k}\cdot \mathbf{x}} + 
	\mathrm{\,c.c.},
\end{subeqnarray}
\returnthesubequation
into (\ref{eqn:HVCStabEquations}$a$--$c$), where $\mathbf{k} = (k_1,k_2,k_3)$ is the 3-dimensional wavenumber. Then, the relation between the complex growth rates $s$ and the wavevector $\mathbf{k}$ can be determined from the eigenvalue problem
\begin{subeqnarray}\label{eqn:eigproblem}
	\gdef\thesubequation{\theequation \mbox{\textit{a}},\textit{b}}
	s\widehat{\theta}	= -k^2\widehat{\theta} + \widehat{u}_3,
	\qquad
	s\widehat{u}_i 		= -\Pran\,k^2\widehat{u}_i + 
	\Ray_b\,\Pran\left(-\dfrac{k_1k_i}{k^2} +\delta_{i1}\right) 
	\widehat{\theta},
\end{subeqnarray}
\returnthesubequation
where $k^2 \equiv |\mathbf{k}|^2$. Equation (\ref{eqn:eigproblem}) has as solutions four eigenvalues $s_1,\ldots,s_4$. Out of them
\begin{equation}
s_{4,V} = -\dfrac{1}{2}k^2(\Pran+1) + \dfrac{1}{2k^2} \left[(k^2)^4(\Pran-1)^2 
- 4k_1k_3k^2\Pran\Ray_b \right]^{1/2} \label{eqn:VC}
\end{equation}
is the growth rate of interest since the real component of $s_{4,V}$ can be positive. Thus, we associate $s_{4,V}$ with an unstable mode. The growth rates for HVCws are labelled with a subscript $V$. When the real component of $s_{4,V}$ is positive, the imaginary part is always zero, whereas when the real component of $s_{4,V}$ is negative, the imaginary part can be either zero or have a finite value. The remaining real component of the three eigenvalues $s_{1}$, $s_{2}$ and $s_{3}$ are always negative, \ie corresponding to stable modes. For homogeneous RBC, we label the growth rates with the subscript ${RB}$. Its largest one, which is similarly associated with an unstable mode, takes the form
\begin{equation}
s_{4,\textit{RB}} = 
- \dfrac{1}{2}k^2(\Pran+1) + \dfrac{1}{2k^2}\left[(k^2)^4(\Pran-1)^2 
+ 4\left(k_2^2+k_3^2\right)k^2\Pran\Ray_b\right]^{1/2}, \label{eqn:SRB}
\end{equation}
\citep[cf.\,][]{Calzavarini+others.2005,Calzavarini+others.2006}. The imaginary component of $s_{4,\textit{RB}}$ is always zero. Similar to HVCws, the remaining three eigenvalues for homogeneous RBC other than (\ref{eqn:SRB}) are always negative. In addition, there are no zero-mode cases in both HVCws and homogeneous RBC since they only occur when $\mathbf{k} = (0,0,0)$ which corresponds to an unbounded infinite system \citep{Schmidt+others.2012}. If the solutions of homogeneous RBC are assumed to be independent of $x_1$, \ie $k_1 = 0$, equation (\ref{eqn:SRB}) simplifies to equation (9) of \cite{Calzavarini+others.2006} for which $s_{4,\textit{RB}}>0$ for $\Ray_b > (k_2^2 + k_3^2)^2$.

A convenient way to interpret the stability of the modes corresponding to $s_4$ is to visualise the surfaces of marginally stable modes which have $s_{4} = 0$. These isosurfaces are plotted in 3-dimensional wavenumber space in figure \ref{fig:3DModes} for both HVCws and homogeneous RBC at $\Ray_b = 2\times10^4$ and $\Pran = 0.709$. A low $\Ray_b$ is chosen for both configurations in order to relate the unstable modes to the behaviour of exponential solutions of $\Nus_b$ that are known to exist in homogeneous RBC \citep{Calzavarini+others.2005,Calzavarini+others.2006}. In figure \ref{fig:3DModes}, only half of the isosurfaces need to be shown thanks to symmetry: for HVCws (figure \ref{fig:3DModes}$a$), only the region $k_2<0$ is shown since the region is plane-symmetrical at $k_2 = 0$; for homogeneous RBC (figure \ref{fig:3DModes}\textit{b}), only the region $k_2<0$ is shown since the region is axisymmetric about $\mathbf{k} = (k_1,0,0)$. The axisymmetric nature of the isosurface $s_{4,\textit{RB}}=0$ reflects the invariance of homogeneous RBC towards rotation in the horizontal $x_2x_3$-plane \citep{Calzavarini+others.2006,Schmidt+others.2012}. In contrast, the isosurface of $s_{4,V}=0$ shows directional dependence in HVCws. The red markers represent the $\mathbf{k}$-locations of eigenvalues and any solid red markers enclosed within the isosurface represent positive eigenvalues associated with unstable modes. In figure \ref{fig:3DModes}, there are 2 unstable modes in HVCws compared to 16 unstable modes in homogeneous RBC. Varying $\Ray_b$ for both configurations does not change the shape of the isosurface and merely scales the sizes of the stability regions. Thus, increasing $\Ray_b$ will result in a larger region which encompasses more unstable modes. Since the shapes of the stability regions are unchanged with $\Ray_b$, it is immediately apparent that the number of positive eigenvalues for homogeneous RBC is larger compared to HVCws at any given $\Ray_b$. This suggests that at any given $\Ray_b$, the solutions for homogeneous RBC are exposed to more interactions between unstable modes as compared to the solutions for HVCws. In \S\,\ref{sec:CompareToDNS}, we will compare the results of our DNS of HVCws and homogeneous RBC at matched $\Ray_b$. We find that the increased interactions between unstable modes in homogeneous RBC correlate with increased unsteadiness in the time-evolution of $\Nus_b(t)$ as compared to HVCws.

\begin{figure}
	\def\drawline#1#2{\raise 2.5pt\vbox{\hrule width #1pt height #2pt}}
	\def\spacce#1{\hskip #1pt}
	\centering
	\centerline{\includegraphics{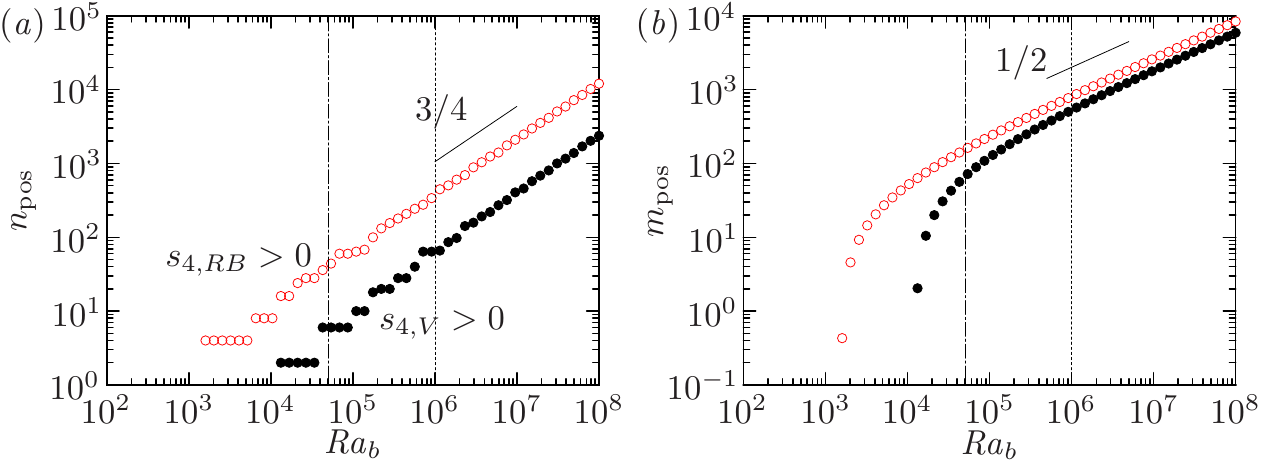}}
	\caption{\label{fig:CountPosModes}($a$) Count of unstable modes $n_{\text{pos}}$ using (\ref{eqn:VC}) for HVCws ({\color{black}$\bullet$}) and (\ref{eqn:SRB}) for homogeneous RBC	({\color{red}$\circ$}). ($b$) Largest magnitude of unstable modes $m_{\text{pos}}$; each point corresponds to the most unstable mode. In ($a$), both trends grow as $\Ray_b^{3/4}$ and at matched $\Ray_b$, $n_{\text{pos},RB} > n_{\text{pos},V}$. In ($b$), the largest magnitudes of the unstable modes vary as $\Ray_b^{1/2}$. The vertical lines represent $\Ray_b \simeq 5\times10^4$		({\color{black}\hbox{\drawline{4}{0.5}\spacce{1}\drawline{1}{0.5}\spacce{1}\drawline{4}{0.5}\spacce{1}\drawline{1}{0.5}}}) and $\Ray_b \simeq 1\times10^6$ 		({\color{black}\hbox{\drawline{1}{0.5}\spacce{1}\drawline{1}{0.5}\spacce{1}\drawline{1}{0.5}\spacce{1}\drawline{1}{0.5}\spacce{1}\drawline{1}{0.5}}}). It can be seen that HVCws has fewer numbers of less unstable modes whereas homogeneous RBC has larger numbers of more unstable modes.}
\end{figure}
To better describe the trend of the unstable modes, the modes for HVCws and homogeneous RBC are quantified for $\Ray_b=10^3$--$10^8$ by considering the number of positive modes (denoted by $n_{\text{pos}}$) for a range of wavenumbers $\mathbf{k} = 2\pi(n_1,n_2,n_3)$ where $(n_1,n_2,n_3)$ are the positive- and negative-integer wavenumbers \citep{Calzavarini+Others+Euromech.2006}. The trend is plotted in figure \ref{fig:CountPosModes}($a$), where black markers represent $n_{\text{pos},V}$ and red markers represent $n_{\text{pos},RB}$. From the figure, we find that $n_{\text{pos},RB}$ is always larger than $n_{\text{pos},V}$, which is consistent with the number of unstable modes counted at matched $\Ray_b$ for the two configurations of figure \ref{fig:3DModes}. This suggests that the solutions for homogeneous RBC will always be more unsteady at all $\Ray_b$ compared to HVCws. We note that both $n_{\text{pos}}$ vary as $\Ray_b^{3/4}$ at high $\Ray_b$. Apart from the positive unstable modes, there exist sets of stable modes (which have $s_4<0$) in the range of $\Ray_b$, not shown in figure \ref{fig:CountPosModes}(\textit{a}). To gain further insight into the stability of the two configurations, the dominance of the most unstable mode at matched $\Ray_b$ is compared in figure \ref{fig:CountPosModes}(\textit{b}) where we plot the magnitudes of the largest positive real component of $s_4$ (denoted by $m_{\text{pos}}$) in the same $\Ray_b$ range over all wavenumbers. From the figure, we observe that $m_{\text{pos}}$ for homogeneous RBC is always larger than that of HVCws (both trends vary as $\Ray^{1/2}$ at high $\Ray_b$).	Thus, from a direct comparison of the trends in figure \ref{fig:CountPosModes}, it could be inferred that the relatively more unsteady solutions for homogeneous RBC (due to higher $n_{\text{pos}}$) will present at a larger magnitude on average (due to higher $m_{\text{pos}}$) as compared to HVCws. In \S\,\ref{sec:CompareToDNS}, we show that the trends from DNS of homogeneous RBC exhibit a higher level of unsteadiness about a larger mean, which is consistent with the observations from the analyses above.

\section{Comparison to DNS} \label{sec:CompareToDNS}
\begin{figure}
	\def\drawline#1#2{\raise 2.5pt\vbox{\hrule width #1pt height #2pt}}
	\def\spacce#1{\hskip #1pt}
	\centering
	\centerline{\includegraphics{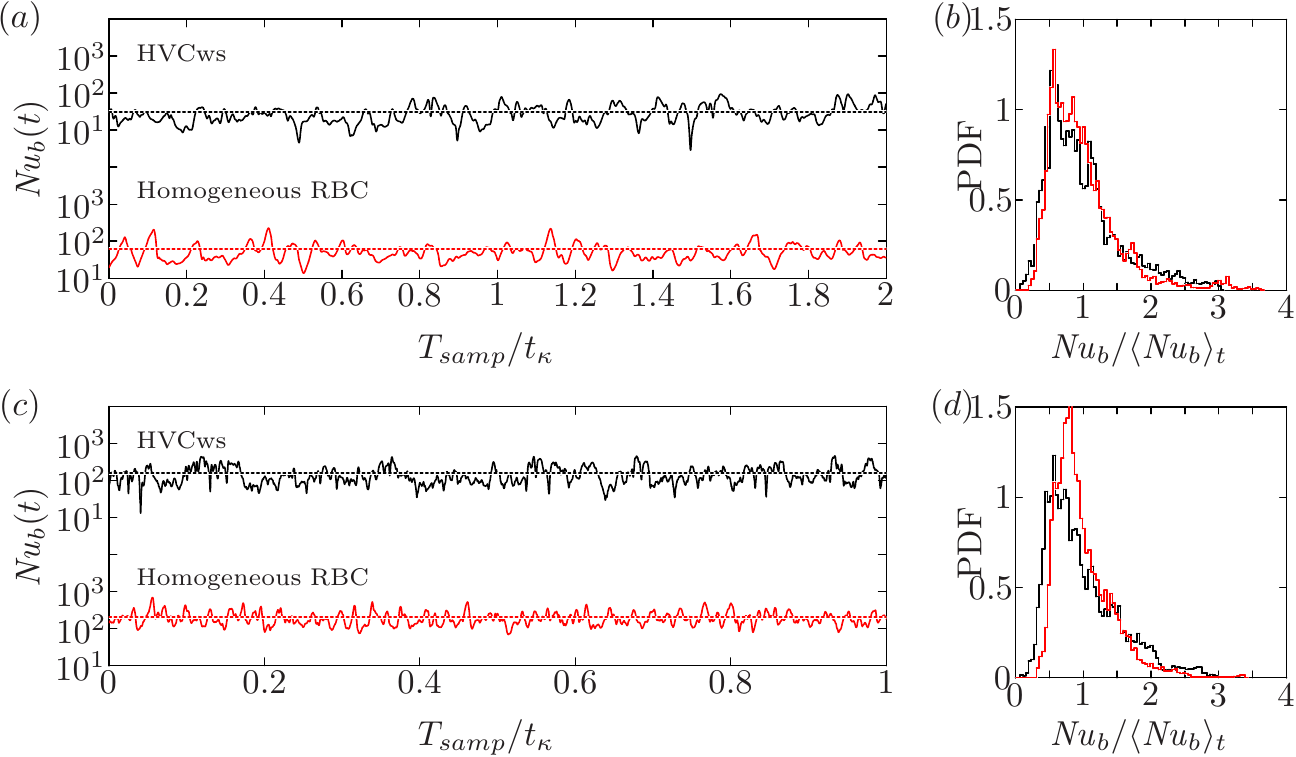}}
	\vspace*{1.5em}
	\caption{\label{fig:NuTime}$\Nus_b(t)$ for HVCws (\hbox{\drawline{8}{0.5}}) and homogeneous RBC ({\color{red}\hbox{\drawline{8}{0.5}}}) for	($a$) $\Ray_b \simeq 5\times10^4$ and ($c$) $\Ray_b \simeq 1\times10^6$ at $\Pran = 0.709$. ($b$,\,$d$) The corresponding PDFs of $\Nus_b(t)$ normalised by the time-averaged $\Nus_b$, indicated by the dotted lines in ($a$) and ($c$). At lower $\Ray_b$ in ($b$), the PDFs for both configurations are positively skewed, which suggest the presence of elevator modes \citep{Calzavarini+others.2005}.	At higher $\Ray_b$ in ($d$), the PDFs for both configurations are less skewed, suggesting that the solutions are subjected to a stabilising mechanism.}
\end{figure}
To further illustrate the results of \S\,\ref{sec:ExponentialGrowth}, we compare the instantaneous Nusselt numbers for HVCws and for homogeneous RBC, and for two relatively low Rayleigh numbers: $\Ray_b \simeq 5\times10^4$ and $1\times10^6$. Both $\Ray_b$-values are indicated respectively by the dot-dashed and dotted lines in figure \ref{fig:CountPosModes}. For homogeneous RBC, we employ the same box size as that for HVCws, \ie\,$L_x = 4L_y = 2L_z$, and the simulations are resolved up to at least $k_\text{max}\eta \approx 2.71$. The simulation parameters for HVCws have been described earlier in \S\,\ref{sec:ComputParam}. We also define the Nusselt number for homogeneous RBC similar to (\ref{eqn:BulkRaReNu}$b$), \ie\,$\Nus_b \equiv JL/(\Delta_b \kappa)$, but with $J \equiv -\kappa \Delta_b/L + \langle u'\varTheta'\rangle$ the vertical heat flux and $L \equiv L_x$.

The instantaneous values of $\Nus_b(t)$ are plotted in figure \ref{fig:NuTime}($a$) for $\Ray_b \simeq 5\times10^4$ and in figure \ref{fig:NuTime}($c$) for $\Ray_b \simeq 1\times10^6$. The corresponding probability distribution functions (PDFs) are shown in figures \ref{fig:NuTime}($b$) and \ref{fig:NuTime}($d$). From the PDFs, the behaviours of $\Nus_b(t)$ of both HVCws and homogeneous RBC appear qualitatively similar at matched $\Ray_b$. Whilst all PDFs at both lower- and higher-$\Ray_b$ appear skewed, both PDFs at the lower-$\Ray_b$ value are relatively more positively skewed, which is consistent with the exponentially growing notion of elevator modes at low $\Ray_b$ and is similar to previously reported PDFs of the time evolution of $\Nus_b(t)$ \citep[see for example figure 7$a$ in][]{Calzavarini+others.2005}. The change from a skewed distribution at lower-$\Ray_b$ to a relatively even distribution at higher-$\Ray_b$ suggests a stabilising mechanism that can be attributed to the presence and interactions of a larger number of unstable modes at higher-$\Ray_b$, as suggested by \cite{Calzavarini+others.2006} and \cite{Schmidt+others.2012}.

When comparing the time-averaged values of $\Nus_b(t)$ for HVCws and homogeneous RBC, we find that the $\Nus_b$ is on average larger for homogeneous RBC than for HVCws (figures \ref{fig:NuTime}$a$,$c$). These trends agree with the presence of larger numbers of larger unstable modes in homogeneous RBC compared to HVCws, as discussed previously in \S\,\ref{sec:ExponentialGrowth}.

\section{Conclusions}
Using a series of DNS of homogeneous vertical natural configuration (HVC -- with shear, HVCws -- without shear) for $\Ray_b$ ranging between $10^5$ and $10^9$ and $\Pran$-value of 0.709, we find that the Nusselt and Reynolds numbers exhibit close to $\Nus_b\sim\Ray_b^{1/2}$ and $\Rey_b\sim\Ray_b^{1/2}$ scaling (figure \ref{fig:NuReRaScaling}), which are consistent with the scaling laws predicted for turbulent bulk-dominated thermal convection at high $\Ray$ \citep{Kraichnan.1962,Grossmann+Lohse.2000}. The present $1/2$-power-law scalings are not only consistent with results from previous studies on homogeneous thermal convection, but can also be found in VC when bulk quantities are employed in the definitions of the Nusselt, Reynolds and Rayleigh numbers.

These $1/2$-power-law scaling results for the homogeneous setups follow in the wake of the scaling arguments for the turbulent bulk flow. In \S\,\ref{sec:UltimateScalingInHVC}, we show that both homogeneous setups are expected to follow the $1/2$-power-law scalings for both Nusselt and Reynolds numbers, consistent with the spirit of the original derivation in \cite{Grossmann+Lohse.2000}. However, in contrast to homogeneous RBC for which the $1/2$-power-law can be conveniently derived \citep[as has been shown in][]{Lohse+Toschi.2003}, the $1/2$-power-law in the homogeneous setup for VC is contingent on several necessary assumptions: the most important assumption being that the vertical turbulent heat flux scales with the horizontal turbulent heat flux and is $\Ray_b$-independent. We show that the $\Ray_b$-independence not only holds true for the homogeneous cases investigated, but also in the bulk region of VC (figure \ref{fig:assumptions}). Other assumptions, whilst exhibiting slight $\Ray_b$-trends, appear to minimally affect the $1/2$-power-law scaling result. 

Although HVC is reminiscent of the bulk of the wall-bounded counterpart (both exhibiting a mean temperature gradient and mean shear), we emphasize that both HVC and HVCws are merely idealisations of the turbulent bulk flow of vertical natural convection. More importantly, the homogeneous cases rightfully capture the orthogonal action of the turbulent vertical convective heat flux and the horizontal heat flux. In addition, the governing equations for the homogeneous cases obey the original scaling arguments in \cite{Grossmann+Lohse.2000} and \cite{Lohse+Toschi.2003} (\ie\,when the contributions of the dissipation in the boundary layers are minimal).

Focussing now on the solutions for HVCws, from stability analysis, we find that the flow is direction-dependent whereas homogeneous RBC is invariant in the vertical direction (figure \ref{fig:3DModes}). In addition, solutions for HVCws are found to be always influenced by fewer numbers of less unstable modes as compared to homogeneous RBC, which are always influenced by larger numbers of more unstable modes relative to HVCws (figure \ref{fig:CountPosModes}). At low to moderate $\Ray_b$, we observe unsteadiness in the solutions (figure \ref{fig:NuTime}), similar to the elevator modes previously reported for homogeneous RBC \citep{Calzavarini+others.2005,Calzavarini+others.2006} and axially homogeneous RBC \citep{Schmidt+others.2012}. In spite of the unsteadiness, results from the DNS appear robust and exhibit close to the $\Ray_b^{1/2}$ power-law scalings for both Nusselt and Reynolds numbers.

The results from the present work suggest that the asymptotic ultimate $1/2$-power-law scaling can also be found in VC in both the homogeneous cases and in the bulk-region of VC. However, the $1/2$-power-law scaling relies on additional assumptions which reasonably agree with the present DNS, but may require further validation in different Rayleigh and Prandtl number regimes.

\section*{Acknowledgements}
This work was supported by the resources from the National Computational 
Infrastructure (NCI) National Facility in Canberra Australia, which is supported by 
the Australian Government, and the Pawsey Supercomputing Centre, which is funded by 
the Australian Government and the Government of Western Australia. DL acknowledges 
support from FOM via the programme``Towards ultimate turbulence''.

\begin{figure}
	\centering
	\centerline{\includegraphics{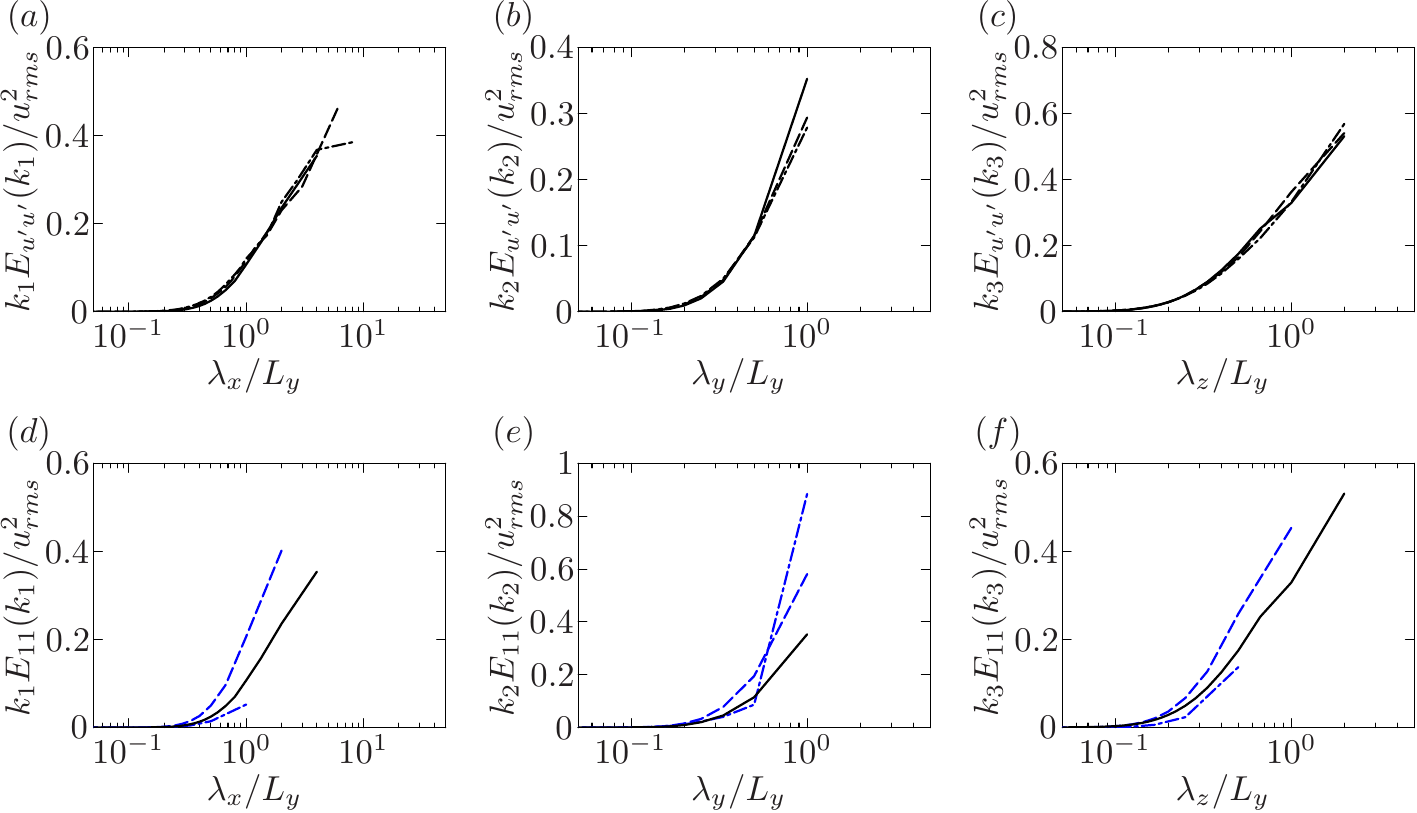}}
	\vspace*{0.5em}
	\caption{\label{fig:TwoPointCorr}Premultiplied one-dimensional spectra of $u'u'$ ($a,d$) in $x$-direction, ($b$,$e$) in $y$-direction, and ($c$,$f$) in $z$-direction. In ($a$--$c$), the streamwise domain is varied where $L_x = 4L_y = 2L_z$ (solid black line, used in our study), $L_x = 6L_y = 3L_z$ (dashed black line) and $L_x = 8L_y = 4L_z$ (dot-dashed black line). In ($d$--$f$), the spanwise domain is varied where $L_x = 2L_y = 2L_z$ (dashed blue line) and $L_x = L_y = 2L_z$ (dot-dashed blue line).}
\end{figure}

\appendix
\section{Sensitivity of box dimensions on homogeneous simulations}
\label{sec:AppendixA}

Here, we provide an analysis on the sensitivity of the box dimensions for the HVC setup. We will make use of the one-dimensional energy spectrum of streamwise velocity $u$, which we define as $E_{u'u'}(k_i)$, where $(u')^2 = 2\int_0^\infty E_{u'u'}\,\romd k_i$, $k_i$ is the wavenumber in the $i$th direction, $\lambda_i = 2\pi k_i^{-1}$ is the corresponding wavelength and $i=1,2$ or $3$. In addition, the spectra is plotted in premultiplied form, which provides an intuitive representation on a logarithmic plot since the area under the curve of a premultiplied spectrum visually represents the distribution of energy that reside at the corresponding  wavelength. The simulation parameters are set such that we match the values of $\Ray_b$ ($=10^5$) and $\Rey_y \equiv SL_y^2/\nu$ ($\approx 640$), which is the box-width Reynolds number \citep{Sekimoto+Dong+Jimenez.2016}. The results are shown in figure \ref{fig:TwoPointCorr}: In ($a$-$c$), $L_x$ is varied while holding $L_y$ and $L_z$ fixed; in ($d$-$f$), $L_y$ is varied while holding $L_x$ and $L_z$ fixed.
	
In figure \ref{fig:TwoPointCorr}($a$-$c$), the spectra collapses when the spanwise domainlength $L_y$ is used to scale the abscissae, suggesting that the limiting domain $L_y$ is indeed the characteristic lengthscale, in agreement with the results of \cite{Sekimoto+Dong+Jimenez.2016}, and is a suitable choice to define $\Ray_b$. In addition, any increase of the streamwise domainlength does little to close the spectra at the longest wavelengths, which suggest that the energetic wavelengths grow to fill the size of the boxes. A similar behaviour is observed in figure \ref{fig:TwoPointCorr}($d$-$f$), where $L_y$ is varied while $L_x$ and $L_z$ are fixed. The latter result indicate that the dynamics of the flow remain sensitive to the box size. As a compromise between computational cost and resolving the large scales in our flow, in our study, we select $L_x = 4L_y = 2L_z$.

\section{Comparison of $\langle\varepsilon_{u'}\rangle$ with 
(\ref{eqn:DissipHVC}$a$) and (\ref{eqn:DissipHVCNoShearStep1}$a$)} 
\label{sec:AppendixB}

\begin{figure}
	\centering
	\centerline{\includegraphics{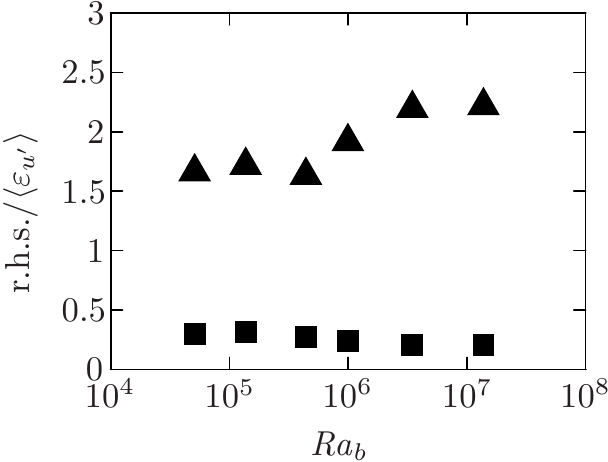}}
	\vspace*{0.5em}
	\caption{\label{fig:EpsRatio}Ratio of r.h.s.\,of (\ref{eqn:DissipHVC}$a$) to $\langle \varepsilon_{u'}\rangle$ for HVC (solid squares) and the r.h.s.\,of (\ref{eqn:DissipHVCNoShearStep1}$a$) to $\langle \varepsilon_{u'}\rangle$ for HVCws (solid triangles). $\langle \varepsilon_{u'}\rangle$ is computed from the gradient of the velocity fluctuations from the DNS.}
\end{figure}

To evaluate the modelled expressions of $\langle \varepsilon_{u'}\rangle$, \ie\,the right-hand-side of (\ref{eqn:DissipHVC}$a$) and the right-hand-side of (\ref{eqn:DissipHVCNoShearStep1}$a$), we show the trend of the ratios with increasing $\Ray_b$ in figure \ref{fig:EpsRatio}. Both trends exhibit weak reliance on $\Ray_b$, albeit slightly more pronounced for the less realistic HVCws setup, which suggest that the models and underlying assumptions for (\ref{eqn:DissipHVC}$a$) and (\ref{eqn:DissipHVCNoShearStep1}$a$) can modestly predict the average kinetic dissipation rate, at least for the present $\Ray_b$-range and $\Pran$-value.

\bibliographystyle{jfm}


\end{document}